\documentclass[conference,10pt,twoside,twocolumn]{IEEEtran}

\usepackage{amsmath,graphicx}

\usepackage[T1]{fontenc}
\usepackage{graphicx}
\usepackage{amssymb}
\usepackage{amsmath}
\usepackage{amsthm}
\usepackage{subfigure}
\usepackage{booktabs} %\toprule \midrule \bottomrule
\usepackage{multirow}
\usepackage{microtype}
\usepackage{cite}
\usepackage{subfigure}
\usepackage{xcolor} % remove xclolor after commenting
\usepackage{url}
\usepackage{balance}
\usepackage{colortbl}
\usepackage{amssymb}
\usepackage{amsfonts}
\usepackage{mathrsfs}
\usepackage{xspace}
\usepackage{bm}
\usepackage{upgreek}

\newcommand{\safemath}[2]{\newcommand{#1}{\ensuremath{#2}\xspace}}

\safemath{\bma}{\mathbf{a}}
\safemath{\bmb}{\mathbf{b}}
\safemath{\bmc}{\mathbf{c}}
\safemath{\bmd}{\mathbf{d}}
\safemath{\bme}{\mathbf{e}}
\safemath{\bmf}{\mathbf{f}}
\safemath{\bmg}{\mathbf{g}}
\safemath{\bmh}{\mathbf{h}}
\safemath{\bmi}{\mathbf{i}}
\safemath{\bmj}{\mathbf{j}}
\safemath{\bmk}{\mathbf{k}}
\safemath{\bml}{\mathbf{l}}
\safemath{\bmm}{\mathbf{m}}
\safemath{\bmn}{\mathbf{n}}
\safemath{\bmo}{\mathbf{o}}
\safemath{\bmp}{\mathbf{p}}
\safemath{\bmq}{\mathbf{q}}
\safemath{\bmr}{\mathbf{r}}
\safemath{\bms}{\mathbf{s}}
\safemath{\bmt}{\mathbf{t}}
\safemath{\bmu}{\mathbf{u}}
\safemath{\bmv}{\mathbf{v}}
\safemath{\bmw}{\mathbf{w}}
\safemath{\bmx}{\mathbf{x}}
\safemath{\bmy}{\mathbf{y}}
\safemath{\bmz}{\mathbf{z}}
\safemath{\bmzero}{\mathbf{0}}
\safemath{\bmone}{\mathbf{1}}
\safemath{\Bell}{\ensuremath{\boldsymbol\ell}}

\bmdefine{\biad}{a}
\bmdefine{\bibd}{b}
\bmdefine{\bicd}{c}
\bmdefine{\bidd}{d}
\bmdefine{\bied}{e}
\bmdefine{\bifd}{f}
\bmdefine{\bigd}{g}
\bmdefine{\bihd}{h}
\bmdefine{\biid}{i}
\bmdefine{\bijd}{j}
\bmdefine{\bikd}{k}
\bmdefine{\bild}{l}
\bmdefine{\bimd}{m}
\bmdefine{\bind}{n}
\bmdefine{\biod}{o}
\bmdefine{\bipd}{p}
\bmdefine{\biqd}{q}
\bmdefine{\bird}{r}
\bmdefine{\bisd}{s}
\bmdefine{\bitd}{t}
\bmdefine{\biud}{u}
\bmdefine{\bivd}{v}
\bmdefine{\biwd}{w}
\bmdefine{\bixd}{x}
\bmdefine{\biyd}{y}
\bmdefine{\bizd}{z}

\bmdefine{\bixid}{\xi}
\bmdefine{\bilambdad}{\lambda}
\bmdefine{\bimud}{\mu}
\bmdefine{\bithetad}{\theta}
\bmdefine{\biphid}{\phi}
\bmdefine{\bideltad}{\delta}

\safemath{\bmia}{\biad}
\safemath{\bmib}{\bibd}
\safemath{\bmic}{\bicd}
\safemath{\bmid}{\bidd}
\safemath{\bmie}{\bied}
\safemath{\bmif}{\bifd}
\safemath{\bmig}{\bigd}
\safemath{\bmih}{\bihd}
\safemath{\bmii}{\biid}
\safemath{\bmij}{\bijd}
\safemath{\bmik}{\bikd}
\safemath{\bmil}{\bild}
\safemath{\bmim}{\bimd}
\safemath{\bmin}{\bind}
\safemath{\bmio}{\biod}
\safemath{\bmip}{\bipd}
\safemath{\bmiq}{\biqd}
\safemath{\bmir}{\bird}
\safemath{\bmis}{\bisd}
\safemath{\bmit}{\bitd}
\safemath{\bmiu}{\biud}
\safemath{\bmiv}{\bivd}
\safemath{\bmiw}{\biwd}
\safemath{\bmix}{\bixd}
\safemath{\bmiy}{\biyd}
\safemath{\bmiz}{\bizd}

\safemath{\bmxi}{\bixid}
\safemath{\bmlambda}{\bilambdad}
\safemath{\bmmu}{\bimud}
\safemath{\bmtheta}{\bithetad}
\safemath{\bmphi}{\biphid}
\safemath{\bmdelta}{\bideltad}

\safemath{\bA}{\mathbf{A}}
\safemath{\bB}{\mathbf{B}}
\safemath{\bC}{\mathbf{C}}
\safemath{\bD}{\mathbf{D}}
\safemath{\bE}{\mathbf{E}}
\safemath{\bF}{\mathbf{F}}
\safemath{\bG}{\mathbf{G}}
\safemath{\bH}{\mathbf{H}}
\safemath{\bI}{\mathbf{I}}
\safemath{\bJ}{\mathbf{J}}
\safemath{\bK}{\mathbf{K}}
\safemath{\bL}{\mathbf{L}}
\safemath{\bM}{\mathbf{M}}
\safemath{\bN}{\mathbf{N}}
\safemath{\bO}{\mathbf{O}}
\safemath{\bP}{\mathbf{P}}
\safemath{\bQ}{\mathbf{Q}}
\safemath{\bR}{\mathbf{R}}
\safemath{\bS}{\mathbf{S}}
\safemath{\bT}{\mathbf{T}}
\safemath{\bU}{\mathbf{U}}
\safemath{\bV}{\mathbf{V}}
\safemath{\bW}{\mathbf{W}}
\safemath{\bX}{\mathbf{X}}
\safemath{\bY}{\mathbf{Y}}
\safemath{\bZ}{\mathbf{Z}}

\safemath{\bZero}{\mathbf{0}}
\safemath{\bOne}{\mathbf{1}}
\safemath{\bDelta}{\mathbf{\Delta}}
\safemath{\bLambda}{\mathbf{\UpLambda}}
\safemath{\bPhi}{\mathbf{\Upphi}}
\safemath{\bSigma}{\mathbf{\Upsigma}}
\safemath{\bOmega}{\mathbf{\Upomega}}
\safemath{\bTheta}{\mathbf{\Uptheta}}

\bmdefine{\biAd}{A}
\bmdefine{\biBd}{B}
\bmdefine{\biCd}{C}
\bmdefine{\biDd}{D}
\bmdefine{\biEd}{E}
\bmdefine{\biFd}{F}
\bmdefine{\biGd}{G}
\bmdefine{\biHd}{H}
\bmdefine{\biId}{I}
\bmdefine{\biJd}{J}
\bmdefine{\biKd}{K}
\bmdefine{\biLd}{L}
\bmdefine{\biMd}{M}
\bmdefine{\biOd}{N}
\bmdefine{\biPd}{O}
\bmdefine{\biQd}{P}
\bmdefine{\biRd}{R}
\bmdefine{\biSd}{S}
\bmdefine{\biTd}{T}
\bmdefine{\biUd}{U}
\bmdefine{\biVd}{V}
\bmdefine{\biWd}{W}
\bmdefine{\biXd}{X}
\bmdefine{\biYd}{Y}
\bmdefine{\biZd}{Z}

\bmdefine{\biDelta}{\Delta}
\bmdefine{\biLambda}{\Lambda}
\bmdefine{\biPhi}{\Phi}
\bmdefine{\biSigma}{\Sigma}
\bmdefine{\biOmega}{\Omega}
\bmdefine{\biTheta}{\Theta}

\safemath{\bimA}{\biAd}
\safemath{\bimB}{\biBd}
\safemath{\bimC}{\biCd}
\safemath{\bimD}{\biDd}
\safemath{\bimE}{\biEd}
\safemath{\bimF}{\biFd}
\safemath{\bimG}{\biGd}
\safemath{\bimH}{\biHd}
\safemath{\bimI}{\biId}
\safemath{\bimJ}{\biJd}
\safemath{\bimK}{\biKd}
\safemath{\bimL}{\biLd}
\safemath{\bimM}{\biMd}
\safemath{\bimN}{\biNd}
\safemath{\bimO}{\biOd}
\safemath{\bimP}{\biPd}
\safemath{\bimQ}{\biQd}
\safemath{\bimR}{\biRd}
\safemath{\bimS}{\biSd}
\safemath{\bimT}{\biTd}
\safemath{\bimU}{\biUd}
\safemath{\bimV}{\biVd}
\safemath{\bimW}{\biWd}
\safemath{\bimX}{\biXd}
\safemath{\bimY}{\biYd}
\safemath{\bimZ}{\biZd}

\safemath{\bimDelta}{\biDelta}
\safemath{\bimLambda}{\biLambda}
\safemath{\bimPhi}{\biPhi}
\safemath{\bimSigma}{\biSigma}
\safemath{\bimOmega}{\biOmega}
\safemath{\bimTheta}{\biTheta}

\safemath{\setA}{\mathcal{A}}
\safemath{\setB}{\mathcal{B}}
\safemath{\setC}{\mathcal{C}}
\safemath{\setD}{\mathcal{D}}
\safemath{\setE}{\mathcal{E}}
\safemath{\setF}{\mathcal{F}}
\safemath{\setG}{\mathcal{G}}
\safemath{\setH}{\mathcal{H}}
\safemath{\setI}{\mathcal{I}}
\safemath{\setJ}{\mathcal{J}}
\safemath{\setK}{\mathcal{K}}
\safemath{\setL}{\mathcal{L}}
\safemath{\setM}{\mathcal{M}}
\safemath{\setN}{\mathcal{N}}
\safemath{\setO}{\mathcal{O}}
\safemath{\setP}{\mathcal{P}}
\safemath{\setQ}{\mathcal{Q}}
\safemath{\setR}{\mathcal{R}}
\safemath{\setS}{\mathcal{S}}
\safemath{\setT}{\mathcal{T}}
\safemath{\setU}{\mathcal{U}}
\safemath{\setV}{\mathcal{V}}
\safemath{\setW}{\mathcal{W}}
\safemath{\setX}{\mathcal{X}}
\safemath{\setY}{\mathcal{Y}}
\safemath{\setZ}{\mathcal{Z}}
\safemath{\emptySet}{\varnothing}

\safemath{\colA}{\mathscr{A}}
\safemath{\colB}{\mathscr{B}}
\safemath{\colC}{\mathscr{C}}
\safemath{\colD}{\mathscr{D}}
\safemath{\colE}{\mathscr{E}}
\safemath{\colF}{\mathscr{F}}
\safemath{\colG}{\mathscr{G}}
\safemath{\colH}{\mathscr{H}}
\safemath{\colI}{\mathscr{I}}
\safemath{\colJ}{\mathscr{J}}
\safemath{\colK}{\mathscr{K}}
\safemath{\colL}{\mathscr{L}}
\safemath{\colM}{\mathscr{M}}
\safemath{\colN}{\mathscr{N}}
\safemath{\colO}{\mathscr{O}}
\safemath{\colP}{\mathscr{P}}
\safemath{\colQ}{\mathscr{Q}}
\safemath{\colR}{\mathscr{R}}
\safemath{\colS}{\mathscr{S}}
\safemath{\colT}{\mathscr{T}}
\safemath{\colU}{\mathscr{U}}
\safemath{\colV}{\mathscr{V}}
\safemath{\colW}{\mathscr{W}}
\safemath{\colX}{\mathscr{X}}
\safemath{\colY}{\mathscr{Y}}
\safemath{\colZ}{\mathscr{Z}}

\safemath{\opA}{\mathbb{A}}
\safemath{\opB}{\mathbb{B}}
\safemath{\opC}{\mathbb{C}}
\safemath{\opD}{\mathbb{D}}
\safemath{\opE}{\mathbb{E}}
\safemath{\opF}{\mathbb{F}}
\safemath{\opG}{\mathbb{G}}
\safemath{\opH}{\mathbb{H}}
\safemath{\opI}{\mathbb{I}}
\safemath{\opJ}{\mathbb{J}}
\safemath{\opK}{\mathbb{K}}
\safemath{\opL}{\mathbb{L}}
\safemath{\opM}{\mathbb{M}}
\safemath{\opN}{\mathbb{N}}
\safemath{\opO}{\mathbb{O}}
\safemath{\opP}{\mathbb{P}}
\safemath{\opQ}{\mathbb{Q}}
\safemath{\opR}{\mathbb{R}}
\safemath{\opS}{\mathbb{S}}
\safemath{\opT}{\mathbb{T}}
\safemath{\opU}{\mathbb{U}}
\safemath{\opV}{\mathbb{V}}
\safemath{\opW}{\mathbb{W}}
\safemath{\opX}{\mathbb{X}}
\safemath{\opY}{\mathbb{Y}}
\safemath{\opZ}{\mathbb{Z}}
\safemath{\opZero}{\mathbb{O}}
\safemath{\identityop}{\opI}

\safemath{\veca}{\bma}
\safemath{\vecb}{\bmb}
\safemath{\vecc}{\bmc}
\safemath{\vecd}{\bmd}
\safemath{\vece}{\bme}
\safemath{\vecf}{\bmf}
\safemath{\vecg}{\bmg}
\safemath{\vech}{\bmh}
\safemath{\veci}{\bmi}
\safemath{\vecj}{\bmj}
\safemath{\veck}{\bmk}
\safemath{\vecl}{\bml}
\safemath{\vecm}{\bmm}
\safemath{\vecn}{\bmn}
\safemath{\veco}{\bmo}
\safemath{\vecp}{\bmp}
\safemath{\vecq}{\bmq}
\safemath{\vecr}{\bmr}
\safemath{\vecs}{\bms}
\safemath{\vect}{\bmt}
\safemath{\vecu}{\bmu}
\safemath{\vecv}{\bmv}
\safemath{\vecw}{\bmw}
\safemath{\vecx}{\bmx}
\safemath{\vecy}{\bmy}
\safemath{\vecz}{\bmz}

\safemath{\veczero}{\bmzero}
\safemath{\vecone}{\bmone}
\safemath{\vecxi}{\bmxi}
\safemath{\veclambda}{\bmlambda}
\safemath{\vecmu}{\bmmu}
\safemath{\vectheta}{\bmtheta}
\safemath{\vecphi}{\bmphi}
\safemath{\vecdelta}{\bmdelta}

\safemath{\matA}{\bA}
\safemath{\matB}{\bB}
\safemath{\matC}{\bC}
\safemath{\matD}{\bD}
\safemath{\matE}{\bE}
\safemath{\matF}{\bF}
\safemath{\matG}{\bG}
\safemath{\matH}{\bH}
\safemath{\matI}{\bI}
\safemath{\matJ}{\bJ}
\safemath{\matK}{\bK}
\safemath{\matL}{\bL}
\safemath{\matM}{\bM}
\safemath{\matN}{\bN}
\safemath{\matO}{\bO}
\safemath{\matP}{\bP}
\safemath{\matQ}{\bQ}
\safemath{\matR}{\bR}
\safemath{\matS}{\bS}
\safemath{\matT}{\bT}
\safemath{\matU}{\bU}
\safemath{\matV}{\bV}
\safemath{\matW}{\bW}
\safemath{\matX}{\bX}
\safemath{\matY}{\bY}
\safemath{\matZ}{\bZ}
\safemath{\matzero}{\bmzero}

\safemath{\matDelta}{\bDelta}
\safemath{\matLambda}{\bLambda}
\safemath{\matPhi}{\bPhi}
\safemath{\matSigma}{\bSigma}
\safemath{\matOmega}{\bOmega}
\safemath{\matTheta}{\bTheta}

\safemath{\matidentity}{\matI}
\safemath{\matone}{\matO}

\safemath{\rnda}{A}
\safemath{\rndb}{B}
\safemath{\rndc}{C}
\safemath{\rndd}{D}
\safemath{\rnde}{E}
\safemath{\rndf}{F}
\safemath{\rndg}{G}
\safemath{\rndh}{H}
\safemath{\rndi}{I}
\safemath{\rndj}{J}
\safemath{\rndk}{K}
\safemath{\rndl}{L}
\safemath{\rndm}{M}
\safemath{\rndn}{N}
\safemath{\rndo}{O}
\safemath{\rndp}{P}
\safemath{\rndq}{Q}
\safemath{\rndr}{R}
\safemath{\rnds}{S}
\safemath{\rndt}{T}
\safemath{\rndu}{U}
\safemath{\rndv}{V}
\safemath{\rndw}{W}
\safemath{\rndx}{X}
\safemath{\rndy}{Y}
\safemath{\rndz}{Z}

\safemath{\rveca}{\bimA}
\safemath{\rvecb}{\bimB}
\safemath{\rvecc}{\bimC}
\safemath{\rvecd}{\bimD}
\safemath{\rvece}{\bimE}
\safemath{\rvecf}{\bimF}
\safemath{\rvecg}{\bimG}
\safemath{\rvech}{\bimH}
\safemath{\rveci}{\bimI}
\safemath{\rvecj}{\bimJ}
\safemath{\rveck}{\bimK}
\safemath{\rvecl}{\bimL}
\safemath{\rvecm}{\bimM}
\safemath{\rvecn}{\bimN}
\safemath{\rveco}{\bomO}
\safemath{\rvecp}{\bimP}
\safemath{\rvecq}{\bimQ}
\safemath{\rvecr}{\bimR}
\safemath{\rvecs}{\bimS}
\safemath{\rvect}{\bimT}
\safemath{\rvecu}{\bimU}
\safemath{\rvecv}{\bimV}
\safemath{\rvecw}{\bimW}
\safemath{\rvecx}{\bimX}
\safemath{\rvecy}{\bimY}
\safemath{\rvecz}{\bimZ}

\safemath{\rvecxi}{\bmxi}
\safemath{\rveclambda}{\bmlambda}
\safemath{\rvecmu}{\bmmu}
\safemath{\rvectheta}{\bmtheta}
\safemath{\rvecphi}{\bmphi}

\safemath{\rmatA}{\bimA}
\safemath{\rmatB}{\bimB}
\safemath{\rmatC}{\bimC}
\safemath{\rmatD}{\bimD}
\safemath{\rmatE}{\bimE}
\safemath{\rmatF}{\bimF}
\safemath{\rmatG}{\bimG}
\safemath{\rmatH}{\bimH}
\safemath{\rmatI}{\bimI}
\safemath{\rmatJ}{\bimJ}
\safemath{\rmatK}{\bimK}
\safemath{\rmatL}{\bimL}
\safemath{\rmatM}{\bimM}
\safemath{\rmatN}{\bimN}
\safemath{\rmatO}{\bimO}
\safemath{\rmatP}{\bimP}
\safemath{\rmatQ}{\bimQ}
\safemath{\rmatR}{\bimR}
\safemath{\rmatS}{\bimS}
\safemath{\rmatT}{\bimT}
\safemath{\rmatU}{\bimU}
\safemath{\rmatV}{\bimV}
\safemath{\rmatW}{\bimW}
\safemath{\rmatX}{\bimX}
\safemath{\rmatY}{\bimY}
\safemath{\rmatZ}{\bimZ}

\safemath{\rmatDelta}{\bimDelta}
\safemath{\rmatLambda}{\bimLambda}
\safemath{\rmatPhi}{\bimPhi}
\safemath{\rmatSigma}{\bimSigma}
\safemath{\rmatOmega}{\bimOmega}
\safemath{\rmatTheta}{\bimTheta}

\usepackage{amssymb}
\usepackage{amsfonts}
\usepackage{mathrsfs}
\usepackage{xspace}
\usepackage{bm}
\usepackage{fancyref}
\usepackage{textcomp}

\usepackage{multirow}
\usepackage{stmaryrd}

\newenvironment{textbmatrix}{	\setlength{\arraycolsep}{2.5pt}%
								\left[\begin{matrix}}{\end{matrix}\right]%
								\raisebox{0.08ex}{\vphantom{M}}}

\def\be{\begin{equation}}
\def\ee{\end{equation}}
\def\een{\nonumber \end{equation}}
\def\mat{\begin{bmatrix}}
\def\emat{\end{bmatrix}}
\def\btm{\begin{textbmatrix}}
\def\etm{\end{textbmatrix}}

\def\ba#1\ea{\begin{align}#1\end{align}}
\def\bas#1\eas{\begin{align*}#1\end{align*}}
\def\bs#1\es{\begin{split}#1\end{split}}
\def\bg#1\eg{\begin{gather}#1\end{gather}}
\def\bml#1\eml{\begin{multline}#1\end{multline}}
\def\bi#1\ei{\begin{itemize}#1\end{itemize}}

\newcommand{\lefto}{\mathopen{}\left}

\DeclareMathOperator{\Prob}{\opP}			% probability of an event
\DeclareMathOperator{\Exop}{\opE}			% expectation operator
\DeclareMathOperator{\Varop}{\opV\!\mathrm{ar}} % variance operator
\DeclareMathOperator{\Covop}{\opC\!\mathrm{ov}}% covariance operator
\newcommand{\Ex}[1]{\ensuremath{\Exop\lefto[#1\right]}} 	% expectation
\newcommand{\Var}[1]{\ensuremath{\Varop\lefto[#1\right]}} % variance
\newcommand{\Cov}[1]{\ensuremath{\Covop\lefto[#1\right]}} % covariance
\newcommand{\abs}[1]{\lefto\lvert#1\right\rvert}		% absolute value

\newcommand{\tp}[1]{\ensuremath{#1^{T}}} 		% transpose
\newcommand{\herm}[1]{\ensuremath{#1^{H}}} 	% hermitian transpose
\newcommand{\inv}[1]{\ensuremath{#1^{-1}}} 	% inverse
\newcommand{\est}[1]{\ensuremath{\hat{#1}}} 	% estimate
\safemath{\dirac}{\delta}					% Dirac delta
\safemath{\krond}{\dirac}					% Kronecker delta
\safemath{\upto}{\uparrow}
\safemath{\downto}{\downarrow}
\safemath{\iu}{j}							% imaginary unit
\safemath{\ev}{\lambda}						% eigenvalue
\safemath{\hilseqspace}{l^{2}}				% Hilbert sequence space
\newcommand{\banachfunspace}[1]{\setL^{#1}}	% Banach function space
\safemath{\hilfunspace}{\banachfunspace{2}}	% Hilbert function space
\safemath{\SNR}{\textit{SNR}} 				% signal to noise ratio
\safemath{\PAR}{\textit{PAR}} 				% signal to noise ratio
\safemath{\No}{N_0}							% noise spectral density
\safemath{\Es}{E_s}							% energy per symbol
\safemath{\Eb}{E_b}							% energy per bit
\safemath{\EbNo}{\frac{\Eb}{\No}}
\safemath{\EsNo}{\frac{\Es}{\No}}

\DeclareMathOperator{\CHop}{\ensuremath{\opH}} % channel operator
\safemath{\tvir}{\rndh_{\CHop}}				% time-varying impulse response
\safemath{\tvtf}{\rndl_{\CHop}}				% 	-''- transfer function
\safemath{\spf}{\rnds_{\CHop}}				% spreading function
\safemath{\bff}{H_{\CHop}}					% bi-freuqency function

\safemath{\ircf}{r_{h}}						% impulse response correlation fn.
\safemath{\tftvcf}{r_{s}}					% scattering function
\safemath{\tfcf}{r_{l}}						% time-frequency correlation fn.
\safemath{\bfcf}{r_{H}}						% bi-frequency correlation fn.

\safemath{\tcorr}{c_h}						% time-correlation function
\safemath{\scf}{c_{s}}						% spreading function
\safemath{\tfcorr}{c_{l}}					% transfer-function correlation
\safemath{\fcorr}{c_{H}}						% frequency-correlation function

\safemath{\mi}{I}							% mutual information
\safemath{\capacity}{C}						% capacity

\safemath{\normal}{\mathcal{N}}			% normal distribution
\safemath{\jpg}{\mathcal{CN}}			% jointly proper Gaussian
\safemath{\mchain}{\leftrightarrow}		% Markov chain
\safemath{\dB}{\,\mathrm{dB}}
\safemath{\dBm}{\,\mathrm{dBm}}
\safemath{\Hz}{\,\mathrm{Hz}}
\safemath{\kHz}{\,\mathrm{kHz}}
\safemath{\MHz}{\,\mathrm{MHz}}
\safemath{\GHz}{\,\mathrm{GHz}}
\safemath{\s}{\,\mathrm{s}}
\safemath{\ms}{\,\mathrm{ms}}
\safemath{\mus}{\,\mathrm{\text{\textmu}s}}
\safemath{\ns}{\,\mathrm{ns}}
\safemath{\ps}{\,\mathrm{ps}}
\safemath{\meter}{\,\mathrm{m}}
\safemath{\mm}{\,\mathrm{mm}}
\safemath{\cm}{\,\mathrm{cm}}
\safemath{\m}{\,\mathrm{m}}
\safemath{\W}{\,\mathrm{W}}
\safemath{\mW}{\, \mathrm{mW}}
\safemath{\J}{\,\mathrm{J}}
\safemath{\K}{\,\mathrm{K}}
\safemath{\bit}{\,\mathrm{bit}}
\safemath{\nat}{\,\mathrm{nat}}

\safemath{\define}{\triangleq}			% definition

\safemath{\equivalent}{\sim}
\safemath{\distas}{\sim}					% distributed according to
\safemath{\sdiff}{\Delta}				% symmetric set difference

\safemath{\reals}{\mathbb{R}}
\safemath{\positivereals}{\reals_{+}}
\safemath{\integers}{\mathbb{Z}}
\safemath{\posint}{\integers_{+}}
\safemath{\naturals}{\mathbb{N}}
\safemath{\posnaturals}{\naturals_{+}}
\safemath{\complexset}{\mathbb{C}}
\safemath{\rationals}{\mathbb{Q}}

\newcommand*{\fancyrefapplabelprefix}{app}		% Appendix
\newcommand*{\fancyrefthmlabelprefix}{thm}		% Theorem
\newcommand*{\fancyreflemlabelprefix}{lem}		% Lemma
\newcommand*{\fancyrefcorlabelprefix}{cor}		% Corollary
\newcommand*{\fancyrefdeflabelprefix}{def}		% Definition
\newcommand*{\fancyrefproplabelprefix}{prop}		% Proposition
\newcommand*{\fancyrefexmpllabelprefix}{exmpl}
\newcommand*{\fancyrefalglabelprefix}{alg}		% Algorithm
\newcommand*{\fancyreftbllabelprefix}{tbl}		% Algorithm

\frefformat{vario}{\fancyrefseclabelprefix}{Sec.~#1}
\frefformat{vario}{\fancyrefthmlabelprefix}{Theorem~#1}
\frefformat{vario}{\fancyreftbllabelprefix}{Tbl.~#1}
\frefformat{vario}{\fancyreflemlabelprefix}{Lemma~#1}
\frefformat{vario}{\fancyrefcorlabelprefix}{Corollary~#1}
\frefformat{vario}{\fancyrefdeflabelprefix}{Definition~#1}
\frefformat{vario}{\fancyreffiglabelprefix}{Fig.~#1}
\frefformat{vario}{\fancyrefapplabelprefix}{Appendix~#1}
\frefformat{vario}{\fancyrefeqlabelprefix}{(#1)}
\frefformat{vario}{\fancyrefproplabelprefix}{Proposition~#1}
\frefformat{vario}{\fancyrefexmpllabelprefix}{Example~#1}
\frefformat{vario}{\fancyrefalglabelprefix}{Algorithm~#1}

\safemath{\dictab}{[\,\dicta\,\,\dictb\,]}

\safemath{\ysig}{\bmy}
\safemath{\ysighat}{\hat{\ysig}}
\safemath{\ysigdim}{M}
\safemath{\xsig}{\bmx}
\safemath{\xsigdim}{N}
\safemath{\nx}{n_x}
\safemath{\zsig}{\bmz}
\safemath{\zsigdim}{\ysigdim}
\safemath{\rsig}{\bmr}
\safemath{\Adict}{\bA}
\safemath{\Adicttilde}{\widetilde{\Adict}}
\safemath{\Adictdim}{\outputdim\times\xsigdim}
\safemath{\avec}{\bma}
\safemath{\avectilde}{\tilde{\avec}}
\safemath{\Bdict}{\bB}
\safemath{\Bdicttilde}{\widetilde{\Bdict}}
\safemath{\Cdict}{\bC}
\safemath{\cvec}{\bmc}
\safemath{\Ddict}{\bD}
\safemath{\Ddictdim}{\ysigdim\times\xsigdim}
\safemath{\dvec}{\bmd}
\safemath{\Ddicttilde}{\widetilde{\bD}}
\safemath{\Bonb}{\bB}
\safemath{\bvec}{\bmb}
\safemath{\Bonbdim}{\ysigdim\times\ysigdim}
\safemath{\noise}{\bmn}
\safemath{\noisedim}{\ysigim}
\safemath{\err}{\bme}
\safemath{\errdim}{\ysigdim}
\safemath{\errset}{\setE}
\safemath{\nerr}{n_e}
\safemath{\delop}{\bP_\errset}
\safemath{\delopc}{\bP_{{\errset}^c}}

\safemath{\cplxi}{\imath}
\safemath{\cplxj}{\jmath}
\safemath{\dict}{\matD}
\safemath{\inputdim}{N}		% number of columns of dictionary D
\safemath{\outputdim}{M}		%number of rows of dictionary D
\safemath{\sparsity}{S}	%sparsity
\safemath{\inputdimA}{{N_a}}	%total number of elements in dictionary A
\safemath{\inputdimB}{{N_b}}	%total number of elements in dictionary B
\safemath{\elemA}{{n_a}}	%number of elements chosen from dictionary A
\safemath{\elemB}{{n_b}}	%number of elements chosen from dictionary B
\safemath{\resA}{\matR_a}	%restriction map to elements of dictionary A
\safemath{\resB}{\matR_b}	%restriction map to elements of dictionary B
\safemath{\subD}{\matS} %subdictionary
\safemath{\subA}{\matS_a} %subdictionary part of A
\safemath{\subB}{\matS_b} %subdictionary part of B
\safemath{\dicta}{\matA} 	% first subdictionary
\safemath{\dictb}{\matB} 	% second subdictionary
\safemath{\hollowS}{H}
\safemath{\hollowA}{H_a}
\safemath{\hollowB}{H_b}
\safemath{\cross}{Z}
\safemath{\coh}{\mu_d}			% coherence dictionary
\safemath{\coha}{\mu_a}			% coherence first subdictionary
\safemath{\cohb}{\mu_b}			% coherence second subdictionary
\safemath{\mubs}{\nu}	%block sub-coherence
\safemath{\cohm}{\mu_m} %mutual coherence
\safemath{\dictset}{\setD}	% set of dictionaries
\safemath{\dictsetp}{\dictset(\coh,\coha,\cohb)}	% set of dictionaries parametrized
\safemath{\dictsetgen}{\dictset_\text{gen}}
\safemath{\dictsetgenp}{\dictsetgen(\coh)}
\safemath{\dictsetonb}{\dictset_\text{onb}}
\safemath{\dictsetonbp}{\dictsetonb(\coh)}

\safemath{\leftside}{U}
\safemath{\rightsideA}{R_a}
\safemath{\rightsideB}{R_b}

\safemath{\indexS}{\setI_S} %set of indices participating in sub-dictionary S

\safemath{\na}{n_a}			% cardinality of set of linearly independent columns of first dictionary
\safemath{\nb}{n_b}			% cardinality of set of linearly independent columns of second dictionary
\safemath{\coeffa}{p_i}	%coefficients for columns of A
\safemath{\coeffb}{q_j}	%coefficients for columns of B
\safemath{\seta}{\setP}		% set of linearly independent columns of A
\safemath{\setb}{\setQ}     % set of linearly independent columns of B
\safemath{\setw}{\setW}	%set of n largest elements of w
\safemath{\setz}{\setZ}	%set of L-n largest elements of z
\safemath{\cola}{\veca}		% generic element of the dictionary A
\safemath{\colb}{\vecb}		% generic element of the dictionary B
\safemath{\cold}{\vecd}		% generic element of the dictionary D
\safemath{\inputvec}{\vecx} 	%coefficient vector (input)
\safemath{\error}{\vece}	%error vector
\safemath{\noiseout}{\vecz} 	%noisy output vector
\safemath{\inputvecel}{x}
\safemath{\inputveca}{\vecx_a}
\safemath{\inputvecb}{\vecx_b}
\safemath{\outputvec}{\vecy}	%output of Dictionary
\safemath{\lambdamin}{\lambda_{\mathrm{min}}}
\safemath{\elltwo}{\ell_2}
\safemath{\ellone}{\ell_1}
\safemath{\ellzero}{\ell_0}
\safemath{\ellinf}{\ell_\infty}
\safemath{\ellinftilde}{\ell_{\widetilde\infty}}
\safemath{\licard}{Z(\coh,\coha,\cohb)}
\safemath{\xsol}{\hat{x}}
\safemath{\xbord}{x_b}		%Solution at the border
\safemath{\xstat}{x_s}		%Solution stationary in l0 prob
\safemath{\xstatLone}{\tilde{x}_s}
\safemath{\order}{\mathcal{O}} %order notation (big O)
\safemath{\scales}{\Theta} %scales as
\safemath{\ones}{\mathbf{1}} %all ones matrix
\safemath{\zeroes}{\mathbf{0}} %all zeroes matrix
\safemath{\thlone}{\kappa(\coh,\cohb)} %treshold l1 problem
\safemath{\constoneA}{\delta} %constant in l1 theorem to save space
\safemath{\constoneB}{\epsilon} %constant in l1 theorem to save space
\safemath{\nlarge}{L}				   %num large elements
\safemath{\sumlarge}{S_\nlarge}
\safemath{\maxlarger}{P_\nlarge}	   % maximum in Gribonval and Nielsen
\safemath{\Pzero}{\textrm{P0}}	
\safemath{\Pone}{\textrm{P1}}
\safemath{\vecfir}{\vecw}			 % \vecv element of the kernel of the dictionary, \vecv=[\vecfir \vecsec]
\safemath{\vecsec}{\vecz}
\safemath{\elvecfir}{w}              % element of vecfir
\safemath{\elvecsec}{z}				 % element of vecsec
\safemath{\nlargefir}{n}
\safemath{\normout}{\gamma}
\safemath{\auxfun}{h}
\safemath{\supp}{\textrm{supp}}%support

\safemath{\indexa}{\ell}
\safemath{\indexb}{r}
\safemath{\indexc}{i}
\safemath{\indexd}{j}

\safemath{\project}{P}%projector

\usepackage{framed}

\IEEEoverridecommandlockouts
\allowdisplaybreaks

\begin{document}

\title{DUIDD: Deep-Unfolded Interleaved Detection \\ and Decoding for MIMO Wireless Systems}

\author{Reinhard Wiesmayr$^\text{1}$, Chris Dick$^\text{2}$, Jakob Hoydis$^\text{2}$, and Christoph Studer$^\text{1}$\\[0.3cm]
\textit{$^\text{1}$ETH Zurich, Switzerland; wiesmayr@iis.ee.ethz.ch and studer@ethz.ch}\\
\textit{$^\text{2}$NVIDIA Corporation; cdick@nvidia.com and jhoydis@nvidia.com}
\thanks{The work of RW and CS was supported in part by an ETH Research~Grant. The work of CS was supported in part by ComSenTer, one of six centers in JUMP, a SRC program sponsored by DARPA. The work of CS was supported in part by the U.S.\ NSF under grants CNS-1717559 and ECCS-1824379.}
\thanks{The authors thank Gian Marti for discussions and comments on this paper.}
\thanks{All code and simulation scripts to reproduce the results of this paper are available on GitHub: {https://github.com/IIP-Group/DUIDD}}
}

\maketitle
% !TEX root = ../22ASILOMAR_DUIDD.tex
% DO NOT REMOVE THE ABOVE COMMENT!

\begin{abstract}
Iterative detection and decoding (IDD) is known to achieve near-capacity performance in multi-antenna wireless systems. We propose deep-unfolded interleaved detection and decoding (DUIDD), a new paradigm that reduces the complexity of IDD while achieving even lower error rates. DUIDD interleaves the inner stages of the data detector and channel decoder, which expedites convergence and reduces complexity. Furthermore, DUIDD applies deep unfolding to automatically optimize algorithmic hyperparameters, soft-information exchange, message damping, and state forwarding. We demonstrate the efficacy of DUIDD using NVIDIA's Sionna link-level simulator in a 5G-near multi-user MIMO-OFDM wireless system with a novel low-complexity soft-input soft-output data detector, an optimized low-density parity-check decoder, and channel vectors from a commercial ray-tracer. Our results show that DUIDD outperforms classical IDD both in terms of block error rate and computational complexity.
\end{abstract}

% !TEX root = ../22ASILOMAR_DUIDD.tex
% DO NOT REMOVE THE ABOVE COMMENT!

\section{Introduction}

With the development of recent and future communication standards, such as 5G \cite{ETSI5G} and 6G \cite{Hoydis2021}, basestations (BSs) are required to process data at higher rates from a larger number of antennas and users while achieving higher spectral efficiency and fulfilling stricter quality-of-service requirements. 
In order to keep up with these trends, the design of novel low-complexity multiple-input multiple-output (MIMO) data detection and channel decoding algorithms is necessary.

Wireless receivers for massive multi-user (MU) MIMO systems typically resort to linear miminum mean-square error (LMMSE) data detection followed by a separate channel decoding stage~\cite{Seethaler2004, Burg2006,Ketonen2010,WYWDCS2014,Lu2014,Yang2015,Peng2018}.
Albeit practical to implement, such conventional pipelines that separate data detection from channel decoding are known to operate far from the fundamental capacity limits.
In contrast, iterative detection and decoding (IDD)  is known to achieve near-capacity performance in MIMO wireless systems \cite{HT2003}. However, the complexity of IDD is already high in conventional, small-scale MIMO systems~\cite{Studer2009diss,Preyss2012,Sun2015}, which likely prevents a deployment in multi-antenna wireless systems that must support more antennas, more simultaneously transmitting users (and possibly more parallel data streams), and higher bandwidths~\cite{WDC2014}. 

\subsection{Contributions}

We propose deep-unfolded interleaved detection and decoding (DUIDD), a novel receiver pipeline for traditional, small scale MIMO as well as massive MU-MIMO wireless systems that surpasses the error rate performance of IDD at  lower computational complexity.
DUIDD interleaves the inner iterations of the soft-input soft-output (SISO) data detection and channel decoding stages, which expedites convergence, thereby reducing the number of inner and outer iterations. In order to automatically optimize algorithm hyperparameters, soft-information exchange, message damping, and state forwarding, DUIDD combines deep unfolding \cite{BalatsoukasStimming2019} with emerging machine learning (ML) frameworks \cite{Hoydis2022} and block error rate (BLER)-based training \cite{Wiesmayr2022a}.
Besides utilizing the SISO minimum mean-square-error (MMSE) parallel interference cancellation (PIC) data detector~\cite{Tuchler2002,asicmimo}, we also propose a novel low-complexity SISO data detector that further reduces complexity.
For channel decoding, we consider a  low-density parity-check (LDPC) decoder with message damping~\cite{Nachmani2018,Lian2018}.
We demonstrate the effectiveness of DUIDD by comparing the BLER performance and computational complexity of classical IDD and DUIDD in a 5G-near MIMO orthogonal frequency-division multiplexing (OFDM) system simulated with NVIDIA's Sionna~\cite{Hoydis2022}.

\subsection{Related Work}
The importance of artificial intelligence (AI) and ML in wireless communications is growing quickly and already an integral part of 5G Release 18~\cite{Lin2022}. AI and ML are expected to play and even more central role in 6G systems~\cite{Hoydis2021} and research is heavily focusing  on ML-based baseband processing, ranging from atomistic (separate and independent) optimizations of neural network (NN)-based MIMO data detectors \cite{OShea2017,Samuel2019,Khani2020} or channel decoders \cite{Cammerer2022},
to fully NN-based transmitters and receivers (e.g., end-to-end learning methods~\cite{Doerner2018, Aoudia2019, Song2022, Cammerer2020}),
and model-driven deep unfolding approaches \cite{Nachmani2018,Lian2018,Zhang2022}. 
However, existing NN-based data detectors
turn out to require orders of magnitude higher complexity compared to classical algorithms that were designed by hand. For example, even when excluding training, the NN-based data detectors
compared in~\cite[Tbl.~I]{Nguyen2022} require
$5\times$ to more than $100\times$ higher runtime than a standard LMMSE data detector.
In contrast, model-driven methods which utilize deep unfolding and ML for automated parameter tuning enable superior error rate performance without increasing the computational complexity.

For the same reason,
 deep unfolding was utilized to optimize either classical data detection or channel decoding algorithms by hyperparameter tuning, or augment classical algorithms with a hypernetwork.
Both methods were applied in \cite{Zhang2022} for improving a
classical expectation propagation (EP) data detector in a deep-unfolded IDD receiver.
Other examples for training a traditional LDPC decoder include message damping and weighted belief-propagation decoding~\cite{Nachmani2018,Lian2018}.
However, to the best of our knowledge, there exists no prior work that jointly optimizes (iterative) detection and decoding pipelines on all the system and component levels, which is part of what we do with the proposed DUIDD paradigm.

% !TEX root = ../22ASILOMAR_DUIDD.tex
% DO NOT REMOVE THE ABOVE COMMENT!

\subsection{Notation}

Column vectors are written in bold lowercase, matrices in bold uppercase, and sets in calligraphic letters.
The superscript~$\inv{}$ denotes the matrix inverse, $^{-\frac{1}{2}}$ the inverse matrix square root, and $\herm{}$ the Hermitian.
In general, we use subscripts for indexing (e.g., $x_{u}$ refers to user equipment $u$), but denote the $K$-dimensional all-zero vector by $\veczero_K$,  the $M\times M$ identity matrix is $\matI_M$, the $M\times K$ all-ones matrix is $\vecone_{M\times K}$, and the $K$-dimensional all-ones vector $\vecone_K$; we omit the dimensions whenever they are obvious in the context.
We refer to the $k$th entry of the $u$th vector $\vecx_u$ by $[\vecx_{u}]_{k}$, to the $m$th row of a matrix~$\matX$ by $[\matX]_{m,:}$, and to the element in the $m$th row and $n$th column by $[\matX]_{m,n}$.
The diagonal matrix $\mathrm{diag}(\vecx)$ contains the elements of $\vecx$ on the main diagonal.
The operator
$\Prob[\cdot]$ denotes probability, $\Ex{\cdot}$ expectation, $\Var{\cdot}$ variance, and $\Cov{\cdot}$ covariance.
We use $\mathcal{NC}\left(\bZero, \matC\right)$ to refer to circularly-symmetric complex Gaussian distribution with covariance matrix $\matC$.

% !TEX root = ../22ASILOMAR_DUIDD.tex
% DO NOT REMOVE THE ABOVE COMMENT!

\section{System Model}

We consider a coded MU-MIMO-OFDM wireless uplink system in which $U$ single-antenna user equipments (UEs) simultaneously transmit data to one $B$-antenna BS.
We now detail the encoder and the system model.

\subsection{OFDM-Based MU Data Transmission} \label{subsec:sys_model_tx}

Within an OFDM frame, each UE $u\in\{1,\dots,U\}$ transmits~$D$ data bits $\vecd_u\in\{0,1\}^D$, which are separately (for each UE) encoded  to $K$ coded bits following the encoding rule $\vecb_u=\mathrm{enc}(\vecd_u)\in\{0,1\}^K$ with code rate $R=\frac{D}{K}$. The encoded bits are mapped onto the zero-mean unit-variance transmit constellation points from the set $\mathcal{Q}$ (e.g., QPSK), to data vectors $\vecs_u\in\mathcal{Q}^V$, where each symbol carries $Q=\log_2|\mathcal{Q}|$ bits and $V=\frac{K}{Q}$. The $V$ symbols are mapped onto the OFDM resource grid with $W$ data-carrying sub-carriers and~$T$ OFDM  symbols, which are divided into $T_D$ data symbols and $T_P = T-T_D$ pilots (used for channel estimation); we have  $V=WT_D$.
Thus, within each OFDM frame (comprising of $WT$ resource elements), all UEs simultaneously transmit one codeword $\vecb_u$ and pilots for channel estimation over the wireless channel.

\subsection{MU-MIMO-OFDM System Model} \label{subsec:mu_mimo_ofdm_rf_channel}
For OFDM resource element (RE) $v\in \{1,\dots,WT\}$, we model the frequency-flat baseband input-output relation as 
\be \label{eq:bb_signal_model}
\vecy_{v} = \matH_{v} \vecs_{v} + \vecn_{v},
\ee
where  $\vecy_{v} \in \mathbb{C}^{B}$ is the BS-side receive vector,
$\matH_{v}\in\mathbb{C}^{B \times U}$ is the channel matrix, 
$\vecs_{v}\in\mathcal{Q}^U$ the vector containing the UEs' transmit symbols within RE $v$, with $[\vecs_{v}]_{u}$ corresponding to the symbol of UE $u$,
and $\vecn_v\in\mathbb{C}^{B}$ models thermal noise with distribution $\mathcal{NC}(\veczero,\mathrm{N_0}\matI)$. 
In what follows,  we omit the RE index $v$ to simplify notation.

Since practical systems must estimate the channel matrix~$\bH$, we model the \emph{estimated} channel matrix as $\widehat{\matH} = \matH - \matE$ with estimation error matrix $\matE$.
This model enables us to rewrite the input-output relation in~\eqref{eq:bb_signal_model} as follows: 
\be
\vecy = \left(\widehat{\matH} +\matE\right)\!\vecs + \vecn = \widehat{\matH}\vecs + \vecn^\prime.
\ee
Here, $\vecn^\prime=\matE \vecs + \vecn$ is the effective noise vector with covariance 
\be
\matC_{\vecn^\prime} = \Cov{\vecn^\prime} = \Cov{\matE \vecs + \vecn} = \matC_{\matE} + \mathrm{N_0}\matI
\ee
and $\matC_{\matE} = \Cov{\matE\vecs}$ is the channel estimation error covariance, which is assumed to be independent of the realization of $\bms$ and provided by the channel estimator. 
Since the effective noise vector $\bmn'$ is correlated, we design our MIMO data detectors for the whitened system model 
\be
\tilde{\vecy}= \matC_{\vecn^\prime}^{-\frac{1}{2}}\vecy = \matC_{\vecn^\prime}^{-\frac{1}{2}}\left(\widehat{\matH}\vecs + \vecn^\prime\right) = \widetilde{\matH} \vecs + \tilde{\vecn},
\label{eq:bb_signal_model_whitened}
\ee
in which $\widetilde{\matH} = {\matC}^{-\frac{1}{2}}\widehat{\matH}$ is the whitened channel estimate  and $\tilde{\vecn}=\matC_{\vecn^\prime}^{-\frac{1}{2}}\vecn^\prime \sim\mathcal{NC}(\veczero, \matI)$ the whitened noise vector.

% !TEX root = ../22ASILOMAR_DUIDD.tex
% DO NOT REMOVE THE ABOVE COMMENT!

\section{Iterative Detection and Decoding} \label{sec:classical_IDD}

We now summarize a conventional IDD receiver for MIMO wireless systems, which will serve as a starting point for the proposed DUIDD paradigm introduced in \fref{sec:duidd}.

\subsection{Classical IDD Receiver}

\begin{figure}[tp]
  \centering
  \includegraphics[width=.85\linewidth]{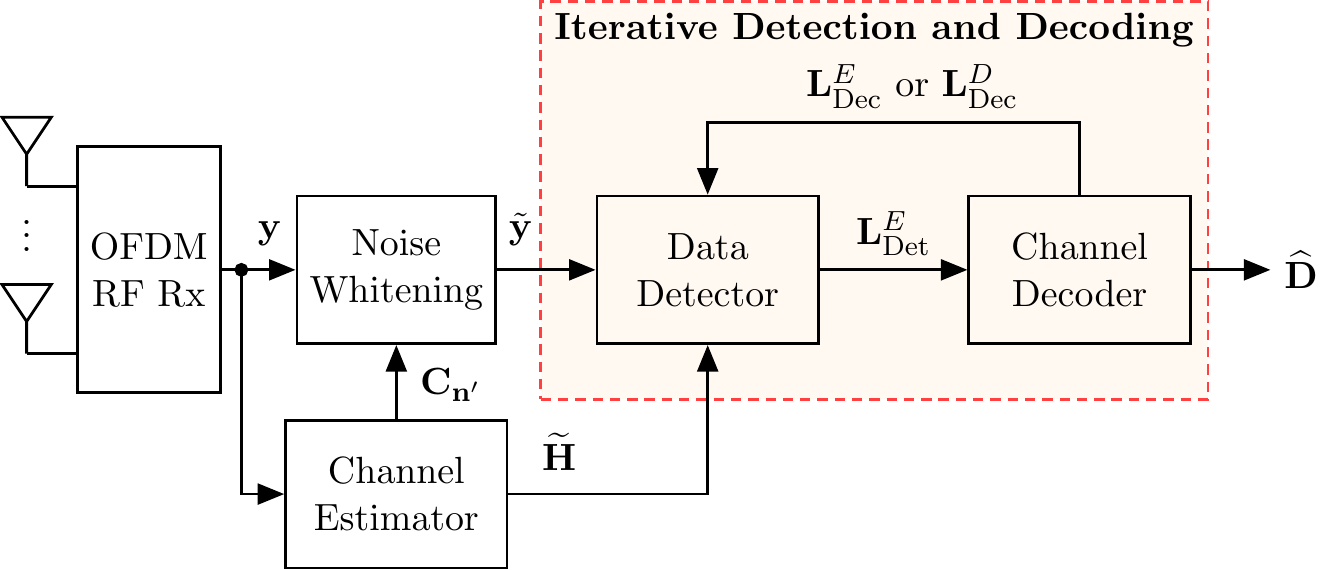}
  \caption{Classical iterative detection and decoding (IDD) MIMO receiver.}
\label{fig:idd_mimo_receiver}
\end{figure}

After receiving an OFDM frame,  the BS first performs channel estimation and noise whitening; see \fref{fig:idd_mimo_receiver} for an illustration. Then, the BS provides the whitened receive vector~$\tilde{\vecy}$ and channel estimate~$\widetilde{\matH}$ to a MIMO data detector, which calculates (approximate) extrinsic log-likelihood ratios (LLR) values $\matL_{\text{Det}}^E\in\mathbb{R}^{U\times K}$ (also known as logits) for the $U$ UEs and $K$ coded bits.
A SISO channel decoder then utilizes these LLR values to calculate (hopefully improved) extrinsic LLR values $\matL_{\text{Dec}}^E=\mathrm{dec}(\matL_{\text{Det}}^E)\in\mathbb{R}^{U\times K}$ for every coded bit\footnote{Since the UEs independently encode their bits $\vecd_{u}$, the decoder \textit{separately} decodes $[\matL_{\text{Dec}}^E]_{u,:}=\mathrm{dec}([\matL_{\text{Det}}^E]_{u,:})\in\mathbb{R}^{K}$ in the considered MU scenario.}.

In a classical IDD receiver (see \fref{fig:idd_mimo_receiver}), the extrinsic LLR values are then fed back to the data detector, where they serve as a-priori LLR values, i.e., $\matL^{A}_{\text{Det}} = \matL^{E}_{\text{Dec}}$ \cite{HT2003}.
A SISO data detector then uses $\tilde{\vecy}$, $\widetilde{\matH}$, and $\matL^{A}_{\text{Det}}$ to compute new (and hopefully improved) extrinsic LLR values that are passed to the SISO channel decoder. The channel decoder produces new LLR values and this process is repeated $I$ times, where $I=1$ means that we use the data detector and channel decoder once.

As outlined above, the data detector takes in a-priori LLR values from the channel decoder, which we define as 
\be \label{eq:a_priori_llr}
[\matL^A_{\text{Det}}]_{u,k} = \log\left(\frac{\Prob[b_{u,k}=1]}{\Prob[b_{u,k}=0]}\right)
\ee
for code bit $k$ of UE $u$.
The data detector then utilizes the whitened receive vector and channel matrix to compute soft-outputs.
By considering a-priori LLR information, the SISO data detector typically first computes intrinsic LLR values that model the a-posteriori bit probabilities, i.e.,
\be \label{eq:intrinsic_llrs}
[\matL^D_{\text{Det}}]_{u,k} = \log\!\left(\frac{\Prob[b_{u,k}=1|\tilde{\vecy},\widetilde{\matH}, \matL^A_{\text{Det}}]}{\Prob[b_{u,k}=0|\tilde{\vecy},\widetilde{\matH},\matL^A_{\text{Det}}]}\right)\!.
\ee

While the intrinsic LLR values are most informative about the current bit-probability belief, the \textit{extrinsic} LLRs model the information gain from the detector, which are usually forwarded to the decoder. By applying Bayes' rule to \eqref{eq:intrinsic_llrs}, the data detector passes \emph{extrinsic}  LLR values to the channel decoder as \cite{HT2003}
\be
\matL^E_{\text{Dec}} = \matL^D_{\text{Dec}} - \matL^A_{\text{Dec}}.
\ee

While exchanging extrinsic LLR values is the de-facto standard,  some data detection algorithms (e.g., the SISO MMSE-PIC algorithm) perform better when \emph{intrinsic} LLR values are fed back from the channel decoder to the data detector~\cite{Witzke2002}, i.e., one directly uses %
$\matL^{A}_{\text{Det}} = \matL^{D}_{\text{Dec}}$. This subtle yet important aspect will be discussed later in \fref{sec:exchange_soft_information}.

After $I$ iterations, the channel decoder for UE $u$ produces hard-output estimates for the transmitted data bits according to $\tp{\hat{\vecd}}_{u} = [\widehat{\matD}]_{u,:}  = \mathrm{u}(\text{dec}^{*}([\matL^{E}_{\text{Det}}]_{u,:}))\in\{0,1\}^{1\times D}$, where $\text{dec}^{*}(\cdot)$ is the soft-output decoding rule for the data bits and $\mathrm{u}$ slices the LLR values to bits in $\{0,1\}$ according to their signs. % italic u is confusing to UE index u

\subsection{SISO MMSE-PIC Algorithm}\label{subsec:siso_mmse_pic}

The SISO MMSE-PIC algorithm is a practical SISO data detector for classical IDD receivers. This algorithm achieves near-capacity performance at manageable complexity and was first described in \cite{Tuchler2002} for systems with inter-symbol interference.
Reference~\cite{asicmimo} reduced the algorithm's  complexity without altering the result, which is the method we detail next.  
For each data-carrying OFDM RE $v\in\{1,\dots,V\}$, the SISO MMSE-PIC algorithm carries out the following four steps where we omit the RE index~$v$ for simplicity.

\subsubsection{Soft-Symbol Computation} 
The algorithm first converts the a-priori LLR values
pertaining to UE $u$ in $\matL^A_{\text{Det}}$
into a-priori soft-symbol estimates $\hat{s}_u^A$ and corresponding variances~$\hat{\nu}^{2,A}_u$.
We define the soft symbol as $\hat{s}_u^A = s_u - e_u^A$, i.e., the true transmit symbol $s_u$ perturbed by an error $e_u^A$, which is assumed to be zero mean with variance~$\hat{\nu}^{2,A}_u$. We compute the soft symbol and variance as follows:
\ba
\hat{s}_u^A &=\textstyle  \sum_{a \in \mathcal{Q}} a \Prob[s_u = a | \matL^A_{\text{Det}}]\\
\hat{\nu}^{2,A}_u &= \textstyle \sum_{a \in \mathcal{Q}} \abs{a-\hat{s}_u^A}^2 \Prob[s_u = a | \matL^A_{\text{Det}}],
\ea
with $\Prob[s_u = a | \matL^A_{\text{Det}}]$ denoting the probability that $s_{u}$ is equal to constellation symbol $a$, which we compute by
\be
\Prob[s_u = a | \matL^A_{\text{Det}}] =\textstyle  \prod_{q=1}^Q \Prob[b_{u,q} = k_{a,q} | [\matL^A_{\text{Det}}]_{u,q}].
\ee
Here, $k_{a,q}$ denotes the $q$th bit of constellation symbol $a$.
We compute the right-hand-side probability as
\be \label{eq:prob_sigmoid}
\Prob[b_{u,q} = k_{a,q} | [\matL^A_{\text{Det}}]_{u,q}] = \sigma\!\left((2 k_{a,q} - 1)[\matL^A_{\text{Det}}]_{u,q}\right)
\ee
with the logistic sigmoid $\sigma(x)=1/(1+e^{-x})$.

\subsubsection{Parallel Interference Cancellation (PIC)} 
Starting from the whitened system model in \eqref{eq:bb_signal_model_whitened}, the algorithm now performs PIC using the computed soft symbols and error terms for UE~$u$ by subtracting the interference of all other users 
\ba
\tilde{\vecy}_u  = \tilde{\vecy} - \!\! \sum_{u^\prime\neq u}^U \tilde{\vech}_{u^\prime}\hat{s}^A_{u^\prime} 
= \tilde{\vech}_{u}s_{u} + \! \sum_{u^\prime\neq u}^U \tilde{\vech}_{u^\prime}e^A_{u^\prime} + \tilde{\vecn}. \label{eq:error_model_after_pic}
\ea

\subsubsection{MMSE Equalization} 
Considering the model after PIC in~\eqref{eq:error_model_after_pic}, we now perform LMMSE equalization to estimate $s_u$. We follow the approach in \cite{asicmimo} and compute
\be \label{eq:mmse_pic_lmmse_matrix_inv}
\inv{\matA} = \big(\herm{\widetilde{\matH}}\widetilde{\matH}\bm{\Lambda} + \matI\big)^{-1},
\ee
with 
$\bm{\Lambda} = \mathrm{diag}\big(\hat{\nu}^{2,A}_1, \dots, \hat{\nu}^{2,A}_U\big)$. The LMMSE equalization  matrix is given by $\matW = \inv{\matA} \herm{\widetilde{\matH}}$, which we use to calculate an \textit{unbiased} estimate for $s_u$ as follows:
\be \label{eq:symbolestimate}
\hat{s}_u^D =  \textstyle \frac{1}{\mu_u}\herm{\vecw_u}\tilde{\vecy}_u.
\ee
Here, $\herm{\vecw_u}$ is the $u$th row of $\matW$ and $\mu_u = \herm{\vecw_u}\tilde{\vech}_u$ is the bias.  As shown in \cite[App.~C]{asicmimo}, the  post-equalization noise-plus-interference (NPI) variance of $\hat{s}_u^D$ can be computed as 
\be
\hat{\nu}_u^{2,D} = \textstyle \frac{1}{\mu_u} - \hat{\nu}_u^{2,A}.
\ee

\subsubsection{LLR Calculation} 
We now use the symbol estimate in~\fref{eq:symbolestimate} to calculate extrinsic max-log LLR values as~\cite[Sec.~III-B]{asicmimo}
\be
[\matL^E_{\text{Det}}]_{u,q} = \frac{1}{\hat{\nu}_u^{2,D}}\Big(\min_{a\in\mathcal{Q}_q^{0}}\abs{\hat{s}_u^D-a}^2 - \min_{a\in\mathcal{Q}_q^{1}}\abs{\hat{s}_u^D-a}^2 \Big),
\ee
where $\mathcal{Q}_q^{b}$ denotes the set of constellation symbols in which the $q$th bit is equal to $b$.

\subsection{LDPC Message Passing Decoding}\label{subsec:siso_ldpc_mp_decoder}
Throughout this paper, we focus on 5G LDPC forward error-correcting codes \cite{ETSI5G} with message passing (MP) decoding. We note that IDD and DUIDD can, in principle, be used together with any other SISO channel decoder. 
The SISO LDPC decoder takes in extrinsic LLR values $\matL_{\text{Det}}^{E}$
from the data detector, performs MP for $N_i$ inner iterations, where $i=1,\ldots,I$ is the outer IDD iteration index, and computes posterior LLR values~$\matL_{\text{Dec}}^{D}$.
For all $I$ IDD iterations, the number of total MP iterations is $N_{\text{MP}}=\sum_{i=1}^I N_i$.
More details about the decoding procedure and our modifications are provided in \fref{sec:damped_mp_ldpc_decoder}.

% !TEX root = ../22ASILOMAR_DUIDD.tex
% DO NOT REMOVE THE ABOVE COMMENT!

\section{Deep Unfolding and Interleaving of IDD} \label{sec:duidd}

We now introduce the ideas behind DUIDD and then detail methods that further improve the efficacy of this paradigm. 

\subsection{DUIDD: Ideas and Concept}

\begin{figure}[tp]
\centering
%\!\!
\subfigure[Classical IDD pipeline.]{\includegraphics[width=.47\columnwidth]{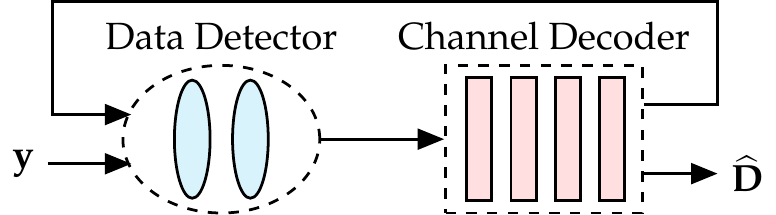}\label{fig:classical_idd}
}
\subfigure[Deep unfolded IDD pipeline.]{\includegraphics[width=.9\columnwidth]{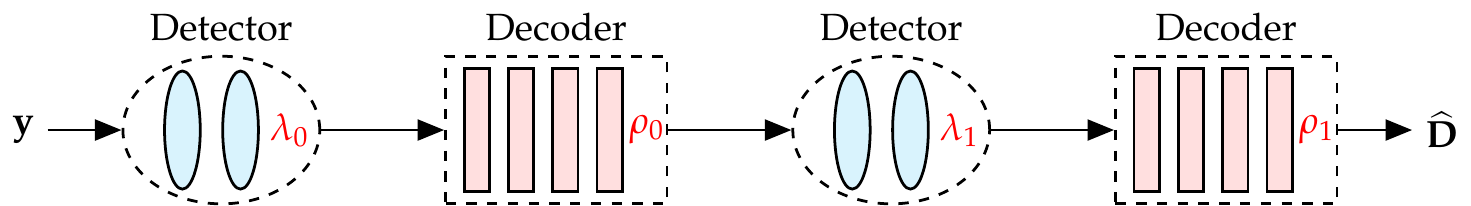}\label{fig:unrolled_deep_unfolded_idd}
}
\subfigure[Deep unfolded interleaved detection and decoding (DUIDD).]{\includegraphics[width=.9\columnwidth]{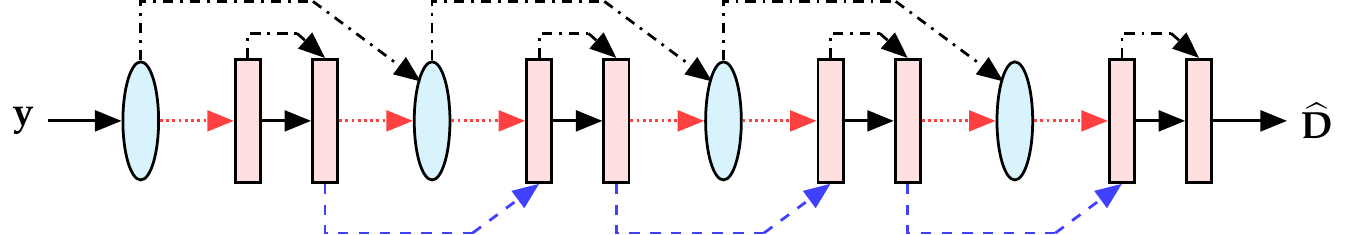}\label{fig:duidd}
}
\caption{Derivation of DUIDD from classical IDD.}
\label{fig:dev_of_duidd}
\end{figure}

DUIDD breaks the strict separation between data detector and channel decoder in a traditional IDD receiver and  jointly optimizes all hyperparameters using deep unfolding and ML to minimize the system's BLER.
The key ideas of DUIDD are illustrated in \fref{fig:dev_of_duidd} and detailed next. 

\subsubsection{Deep-unfolding of IDD}
\fref{fig:classical_idd} shows a classical IDD receiver pipeline with strict separation between data detector and channel decoder. Note that both components may contain inner iterations (shown with blue ellipses and red rectangles), e.g., the MP iterations of a LDPC decoder. 
As shown in \fref{fig:unrolled_deep_unfolded_idd}, DUIDD first applies deep unfolding~\cite{BalatsoukasStimming2019} to the IDD receiver for a fixed number of outer iterations $I$. All hyper-parameters per inner iteration $i=1,\ldots,I$ of the channel decoder and data detector $\lambda_i$ and $\rho_i$, respectively, can then be optimized automatically using ML tools.

\subsubsection{DUIDD}
\begin{figure*}[tp]
\centering
\includegraphics[width=0.97\textwidth]{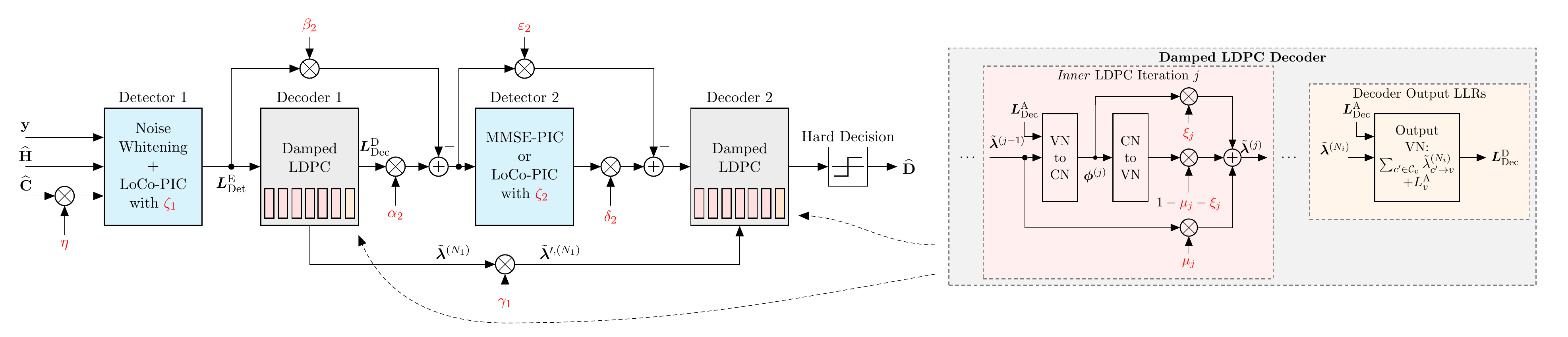}
\vspace{-0.25cm}    
\caption{DUIDD receiver pipeline for $I=2$ unfolded IDD iterations each with $N_{i}=6$ inner LDPC MP decoding iterations.}
\label{fig:duidd_pipeline_I2}
\end{figure*}

As illustrated in \fref{fig:duidd}, the key idea is to break the strict separation of the deep unfolded IDD receiver in \fref{fig:unrolled_deep_unfolded_idd} and interleave (reshuffle) the inner iterations of the data detector and channel decoder. 
The reasons behind interleaving are as follows.
First, interleaving enables an earlier information exchange between the detector and decoder, which can accelerate convergence. Convergence can be optimized by selecting a suitable interleaving pattern, i.e., by specifically deciding on the order between inner detection and decoding stages. 
Second, one can optimize the information exchange across the entire DUIDD pipeline, e.g., by passing extrinsic or intrinsic LLR values (or a mix thereof; see the red dotted arrows), state forwarding (blue dashed arrows), and message damping (black dash-dotted arrows). 
In the ensuing discussion, we apply the DUIDD paradigm to the IDD receiver reviewed in
\fref{sec:classical_IDD} and derive the structure, components, and trainable hyperparameters (highlighted with red) detailed in \fref{fig:duidd_pipeline_I2}.

\subsection{Exchange of Soft-Information}
\label{sec:exchange_soft_information}
In classical IDD, the data detector and channel decoder blocks typically exchange extrinsic LLR values. 
%!TEX encoding = UTF-8 Unicode
In practice, however, some data detectors are known to perform better when fed with intrinsic a-priori LLR values, such as the SISO MMSE-PIC algorithm~\cite{Witzke2002}. 
Furthermore, when using approximate detector or decoder blocks (e.g., max-log turbo decoders), it is known that scaling the extrinsic LLR values can improve convergence of iterative methods \cite{Vogt2000}.
We therefore propose to automatically learn the type and scaling of information exchange by computing the a-priori LLR values for the data detector and channel decoder as follows: 
\ba
\matL^{A}_{\text{Det}} &= \alpha_i \matL^{D}_{\text{Dec}} - \beta_i \matL^{A}_{\text{Dec}}\\ 
\matL^{A}_{\text{Dec}} &= \delta_i \matL^{E}_{\text{Det}} - \varepsilon_i \matL^{A}_{\text{Det}}.
\ea
Here, we introduce new per-outer-iteration
hyperparameters $\left\{\alpha_i, \beta_i, \delta_i, \varepsilon_i\right\}$, $i=1,\ldots,I$. We note that a conventional extrinsic information exchange  is realized by setting $\alpha_i=\beta_i=\delta_i=1$ and $\varepsilon_i=0$, whereas the MMSE-PIC algorithm performs better with $\alpha_i=\delta_i=1$ and $\beta_i=\varepsilon_i=0$. We will use the latter parameters when comparing to classical IDD and to initialize hyperparameter training for DUIDD. 

\subsection{Low-Complexity Soft-Input Soft-Output MIMO Detection}
In a DUIDD receiver, it is beneficial to reduce the complexity of the inner detection and decoding stages as finer granularity can improve the efficacy of interleaving. We now review known methods that reduce complexity of the SISO MMSE-PIC algorithm and then devise a novel, low-complexity variant.

\subsubsection{Low-Complexity SISO MMSE-PIC}
For the Gray-labeling specified in 5G \cite{ETSI5G}, we combine the simplified soft-symbol and variance computation approach from \cite{Tomasoni2006} with the approach in \cite[Tbl.~1]{Collings2004} to directly compute extrinsic max-log LLR values at low complexity. 
Analogous to \cite[Alg.~1]{asicmimo}, we further reduce complexity by rewriting the SISO MMSE-PIC algorithm to operate on the matched filter (MF) output $\tilde{\vecy}^{\mathrm{MF}} = \herm{\widetilde{\matH}}\tilde{\vecy}$.

\subsubsection{SISO LoCo-PIC Algorithm}
Since the computational complexity of the SISO MMSE-PIC algorithm is dominated by computing $\bA^{-1}$ in \fref{eq:mmse_pic_lmmse_matrix_inv} whenever the a-priori LLR values change, it can be advantageous to derive a low-complexity variant.
We propose a low-complexity (LoCo) version of the SISO MMSE-PIC algorithm based on the observation that there exist two corner cases: (i) no useful a-priori information (i.e., all LLR values are zero) and (ii) perfect a-priori information (i.e., the LLR values determine the ground-truth coded bits).

In absence of useful a-priori information, which is generally the case in the first outer iteration, we have $\matL^A_{\text{Det}}=\matzero$ with the resulting soft-symbol estimates $\est{\vecs}^{A}=\veczero$ and error variances~$\est{\bm{\nu}}^{2,A}=\vecone$. 
In this case, we obtain the conventional LMMSE filter matrix  given by~\cite{Seethaler2004} 
\be
\matW_{\mathrm{LM}} = \inv{\left(\herm{\widetilde{\matH}}\widetilde{\matH} + \matI\right)} \herm{\widetilde{\matH}}. 
\ee
With perfect a-priori information,
the soft-symbol estimates correspond to the true transmit symbols and the error variances are zero. 
In this case, we obtain the MF matrix $\matW_{\mathrm{MF}} = \herm{\widetilde{\matH}}$.
The idea of LoCo-PIC is now to model the filter matrix as a linear interpolation between these two corner cases: 
\be \label{eq:locopicfiltermatrix}
\matW_{\mathrm{LC}} =\zeta \matW_{\mathrm{LM}} + (1-\zeta) \matW_{\mathrm{MF}}.
\ee
Here, the parameter $\zeta$ indicates whether we use more or less of the LMMSE filter and MF matrices, respectively. 
While one can derive a closed-form expression for the optimal parameter that minimizes the post-equalization variance when using this filter matrix, evaluating the resulting expression requires high computational complexity. We therefore resort to ML to automatically learn this parameter for every detection stage. 

In order to ensure that we perform unbiased equalization when using \fref{eq:locopicfiltermatrix}, we perform equalization as follows:
\be
\hat{s}_{\text{LC},u}^{D} =\textstyle  \left(\frac{\zeta}{\mu_{\text{LM},u}}{{\vecw_{\text{LM},u}}} + \frac{1-\zeta}{\mu_{\text{MF},u}}{\vecw_{\text{MF},u}}\right)^{\!H}\tilde{\vecy}_u, \label{eq:loco_pic_equalization}
\ee
with bias $\mu_{\text{LM},u} = \herm{\vecw_{\text{LM},u}}\tilde{\vech}_{u}$ and $\mu_{\text{MF},u} =  \| \tilde{\vech}_u\|^2$ and $\herm{\vecw_{\text{LM},u}}$, $\herm{\vecw_{\text{MF},u}}$ denote the $u$th row of $\matW_{\mathrm{LM}}$, $\matW_{\mathrm{MF}}$, respectively.

Since evaluating the closed-form expression for the post-equalization variance of $\hat{s}_{\text{LC},u}^{D}$ requires high complexity,
we take---as a heuristic---the variance of the matched filter equalizer output $\hat{s}_{\text{MF},u}^{D}$ obtained by setting $\zeta=0$ in \eqref{eq:loco_pic_equalization} for the LLR calculation.
In hindsight that the optimal $\zeta$ is expected to be close to 0 for reliable a-priori LLRs, we argue that the NPI variance of $\hat{s}_{\text{MF},u}^{D}$ is a good approximation for the variance of $\hat{s}_{\text{LC},u}^{D}$.
Considering the error model after PIC in \eqref{eq:error_model_after_pic}, we compute the MF NPI variance as 
\be
\hat{\nu}_{\text{MF},u}^{2, D} = \textstyle \sum_{u^\prime\neq u}^U \abs{\frac{[\tilde{\matG}]_{u,u^\prime}}{[\tilde{\matG}]_{u,u}}}^2\hat{\nu}_{u^\prime}^2 + \frac{1}{[\tilde{\matG}]_{u,u}},
\ee
where $\widetilde{\matG}$ denotes the Gram matrix $\widetilde{\matG}=\herm{\widetilde{\matH}}\widetilde{\matH}$.

Note that when we first use LoCo-PIC, which is in absence of any useful a-priori information, we apply the error variance of the conventional LMMSE filter $\nu_{\text{LM},u}^{2,D}={1}/{\mu_{\text{LM},u}} - 1$,
as the optimal parameter $\zeta_{1}$ is expected to be close to $1$.

We conclude by noting that LoCo-PIC has significantly lower complexity than the conventional MMSE-PIC algorithm, as it reuses the same LMMSE and MF filter matrices for all iterations and during the entire coherence interval.  

\subsection{Damped LDPC Message Passing Decoder} 
\label{sec:damped_mp_ldpc_decoder}

We utilize a SISO LDPC decoder that uses the flooding schedule \cite{MacKay1999}. %
To improve performance, we utilize message damping for the check node (CN) to variable node (VN) messages ${\lambda}_{c\to v}$ from CN $c$ to VN $v$.
Since message damping was originally proposed to overcome issues with trapping sets and loopy factor graphs \cite{Savin2021, Nachmani2018},
which become problematic after some MP iterations, we 
modify the damping approaches from \cite{Nachmani2018, Lian2018}
as follows. 
Instead of applying one single \textit{global} damping parameter for all MP iterations, we propose \textit{individual} message damping. We train individual damping~parameters $\mu_j\in[0,1]$ for each MP iteration $j\in\{1,\dots, N_{\text{MP}}\}$.
Additionally, we also include the CN update to message damping.
As shown in the right part of \fref{fig:duidd_pipeline_I2}, we implement the update~rule
\be
\tilde{{\lambda}}_{c\to v}^{(j)} = (1-\mu_j-\xi_j){\lambda}_{c\to v}^{(j)} + \mu_j \tilde{\lambda}_{c\to v}^{(j-1)} + \xi_j {\phi}_{v\to c}^{(j)}
\ee
with $\xi_{j}\in[0,1]$ and the VN to CN message ${\phi}_{v\to c}^{(j)}$, which we compute as follows:
\be
{\phi}_{v\to c}^{(j)} = L^A_v + \textstyle \sum_{c^\prime \in \mathcal{C}_v\backslash c} \tilde{\lambda}_{c^\prime \to v}^{(i-1)}.
\ee
Here,  $\mathcal{C}_v$ specifies the set of check nodes connected to variable node $c$, and $L^A_v$ the a-priori LLR for VN $v$.

The damped MP decoder outputs its intrinsic LLRs by marginalizing the damped CN to VN messages as
 \be
L^D_v= L^A_v +\textstyle  \sum_{c^\prime \in \mathcal{C}_v} \tilde{\lambda}_{c^\prime \to v}^{(j)},
\ee
which are forwarded to the detector after $N_{i}$ MP iterations.

Another important, yet often neglected, aspect of channel decoders with internal states within IDD is whether or not to forward the decoder's state from one IDD iteration to another.
For the considered LDPC MP decoder, the CN to VN messages represent the state after each MP iteration $j$, which we denote by~$\tilde{\bm{\lambda}}^{(j)}$.
As detailed in \cite[Sec.~III-A]{Sun2015}, a non-resetting LDPC MP decoder (initialized with the previous state $\tilde{\bm{\lambda}}^{(N_{i-1})}$) performs substantially better than a resetting decoder (initialized with $\matzero$), particularly for a low number of MP iterations $N_{i}$.
In contrast, a resetting decoder can result in better error rate performance when $N_{i}$ is large.
Aiming at optimizing the initialization method, we train hyperparameters~$\gamma_i$ which scale the forwarded decoder state, i.e., 
we initialize decoder $i+1$ with $\tilde{\bm{\lambda}}^{\prime,(N_{i})} = \gamma_i\tilde{\bm{\lambda}}^{(N_{i})}$. For the classical IDD benchmarks, as well as for the initial value for training, we apply $\gamma_i=1$.

\subsection{Block Error Rate (BLER) Training}

In order to tune all hyperparameters, we aim at minimizing the BLER by using the recent approach from \cite{Wiesmayr2022a}.
We modify the last decoder stage to output LLR values for the data bits and convert these into probabilities similar to \eqref{eq:prob_sigmoid}.
We then train the hyperparameters in two stages. First, we minimize the empirical mean of the binary cross entropy (BCE) loss 
\begin{equation}
\textstyle
 l_{\text{BCE}}(d,\hat{p}) = - d \log(\hat{p}) - (1 - d) \log(1-\hat{p}), \label{eq:lbce}
\end{equation}
with probability estimate $\hat{p}$ for the data bit $d$, by averaging over all transmitted bits within a batch (comprising of $M$ OFDM frame transmissions).
We then continue with BLER refinement training by using the normalized LogSumExp loss function~\cite{Wiesmayr2022a}
\begin{align}
%\textstyle
    l_{\mathrm{LSE}}(\vecd,\hat{\vecp}) =  \log\!\left(\sum_{k=1}^D\exp( l_{\text{BCE}}([\vecd]_{k},[\hat{\vecp}]_{k})) \!-\! D \!+\! 1\! \right)\!,
\end{align}
with probability estimates $\hat{\vecp}$ for data bits $\vecd$, and we average over all transmitted data blocks within a batch. %
For each batch sample, the signal-to-noise-ratio (SNR) is drawn uniformly at random from the considered SNR training-range.

% !TEX root = ../22ASILOMAR_DUIDD.tex
% DO NOT REMOVE THE ABOVE COMMENT!

\section{5G Link-level Simulation in NVIDIA Sionna}\label{sec:simulation_results}

We now demonstrate the efficacy of DUIDD in NVIDIA's Sionna 5G link-level simulator~\cite{Hoydis2022}.
We next discuss the simulation settings and then results for two scenarios: (i) Rayleigh fading channels with perfect channel state information (CSI) and (ii) channels from a ray-tracer with estimated CSI.

\subsection{Simulation and Training Parameters}

We simulate a 5G-near MU-MIMO-OFDM uplink system with $100$\,MHz bandwidth operating at a carrier frequency of $3.75$\,GHz.
We consider $U=4$ single-antenna UEs and the following two BS-antenna configurations: (i) $B=4$ BS antennas when simulating Rayleigh-fading channels and (ii) $B=8$ antennas when using the ray-tracing channels.
We use Sionna's  $R=0.5$ rate 5G LDPC code with rate matching and bit-interleaving as described in \cite[Sec 5.4.2.2]{ETSI5G}. Note that we modified the LDPC decoder to include the changes discussed in \fref{sec:damped_mp_ldpc_decoder}. For each Monte-Carlo trial, we randomly generate $D=1'200$ data bits and map the coded bits to a $16$-QAM constellation with Gray labeling. 
The OFDM resource grid consists of $60$ subcarriers ($5$ resource blocks, each with $12$ sub-carriers), $30$\,kHz subcarrier spacing, and $T=14$ OFDM symbols; $T_P=4$ OFDM symbols are used for pilot-based channel estimation.%
\,We use Sionna's\,Kronecker pilot pattern twice: time slots $t=\{3,12\}$ for UE $1$ and $2$, and time slots $t=\{4,13\}$ for UE $3$ and $4$. We use least-squares channel estimation with linear interpolation between subcarriers and time slots.
We assume perfect power control so that the channel vectors are normalized to have unit average energy per RE. 

For BCE-based pre-training and BLER training, we use the Adam optimizer with $2'500$ batches, each consisting of $M=40$ samples.
The initial hyperparameter values correspond to those of classical IDD.
The $E_b/\No$ 
range during training is $[-5,5]$\,dB for Rayleigh fading channels and $[-5,15]$\,dB for ray-tracing channels.
After training, we evaluate the BLER with $10^5$ unseen samples for each SNR value.
Note that we train a single set of hyperparameter values, which we fix after training and use for the entire SNR range during testing.

\subsection{Rayleigh Fading Channels and Perfect CSI}\label{sec:sim_results_rayleigh}

\begin{figure}[tp]
\centering
\subfigure[]{\includegraphics[width=.5175\linewidth]{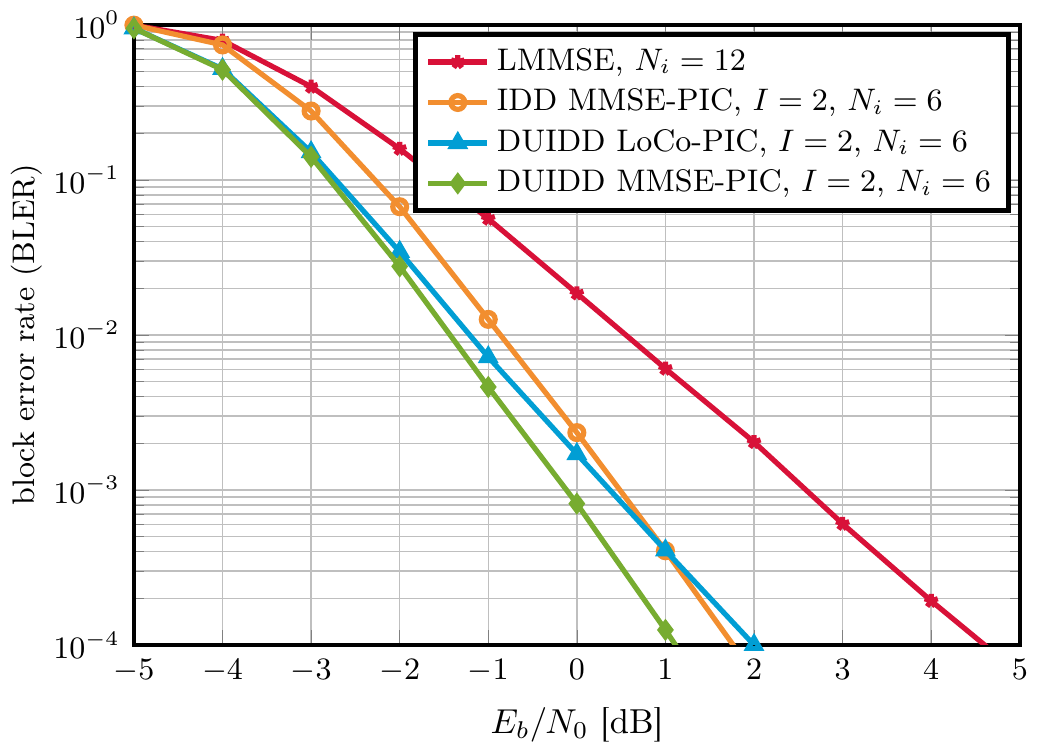}\label{fig:rayleigh_perf_csi_bler_performance}}
\hfill
\subfigure[]{\includegraphics[width=.47\linewidth]{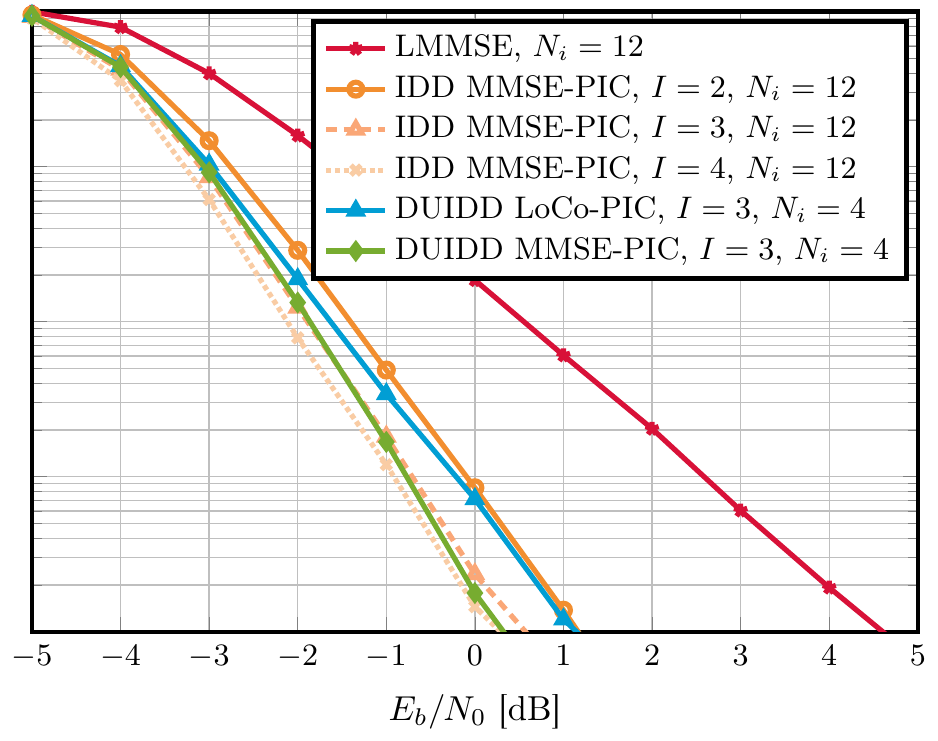}\label{fig:rayleigh_perf_csi_bler_duidd_vs_idd_more_dec_v2}}
\subfigure[]{\includegraphics[width=.5175\linewidth]{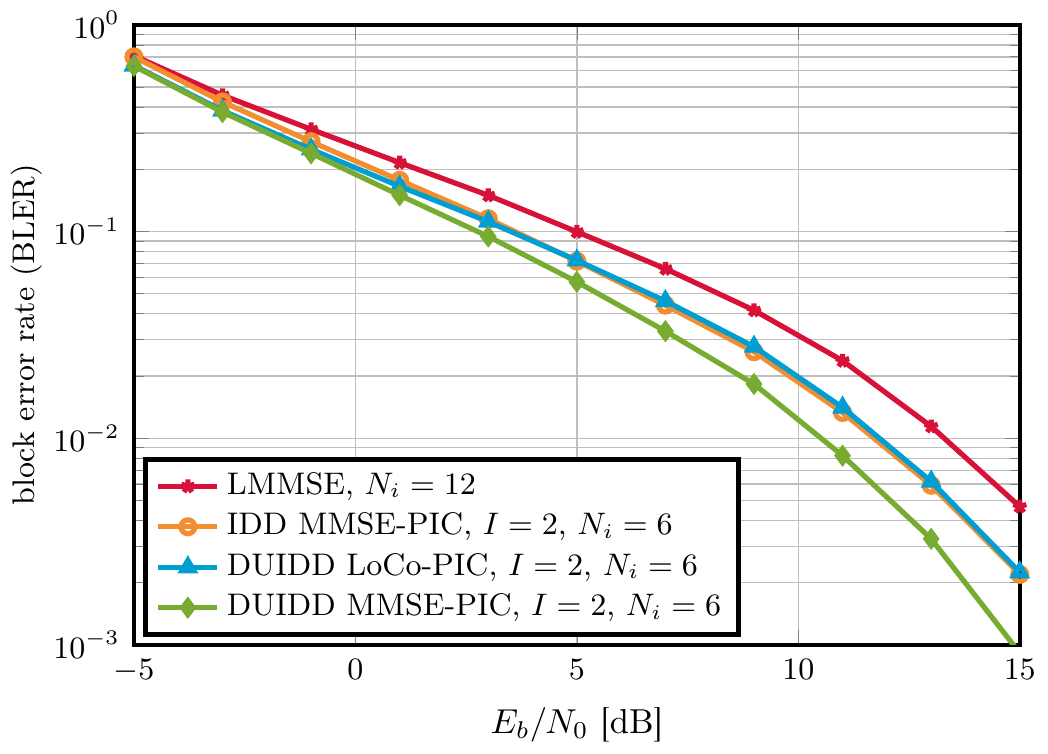}\label{fig:remcom_cest_bler_performance}}
\hfill
\subfigure[]{\includegraphics[width=.47\linewidth]{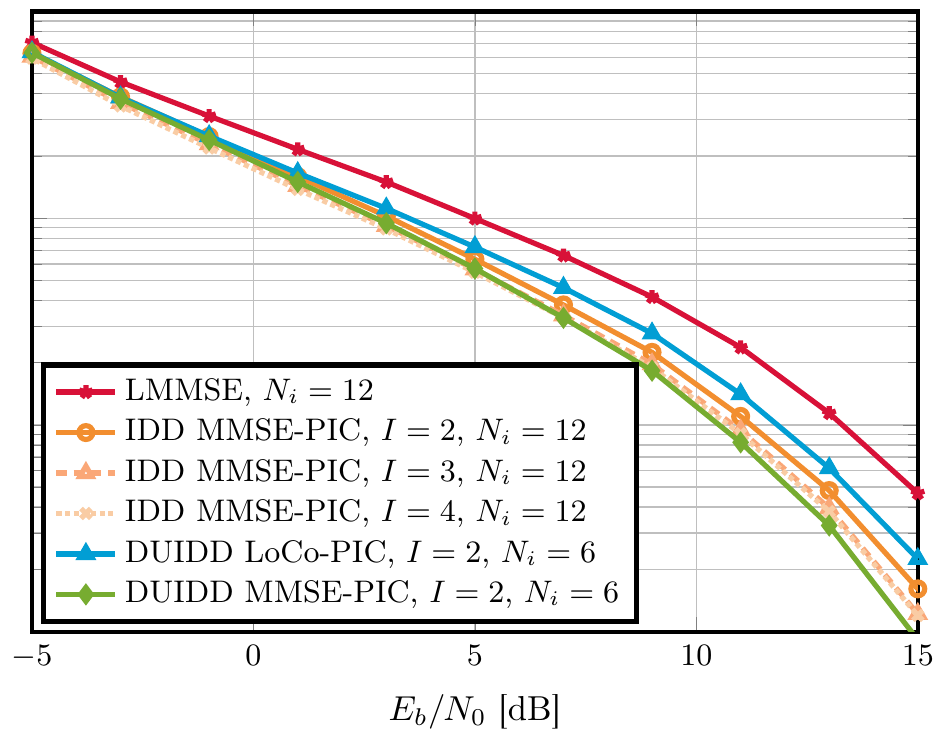}\label{fig:remcom_cest_bler_duidd_vs_idd_more_dec}}
\caption{BLER vs. SNR in a MU-MIMO-OFDM system with $B\times U=8\times4$ Rayleigh fading channels and perfect CSI (top row), or with $B\times U=16\times4$ ray-tracing channels and estimated CSI (bottom row). 
The left column shows results for a fixed total number of $N_{\text{MP}}=12$ LDPC MP decoding iterations ($N_i = \frac{12}{I}$ per detection stage). 
The right column compares DUIDD with $N_{i}={12}/{I}$ LDPC MP decoding iterations per detection stage against IDD with a constant ($N_i=12$) number of iterations per detection stage.}
\end{figure}

\fref{fig:rayleigh_perf_csi_bler_performance} shows simulation results for frequency-flat Rayleigh block fading channel matrices, which are independent and identically distributed for each OFDM block.
Note that we assume perfect CSI. 
We use a total number of $N_{\text{MP}}=12$ LDPC MP decoding iterations, which we evenly interleave with $I\in\{1,\dots,4\}$ detector stages.
Remember that the total number of MP iterations equals the number of LDPC decoding iterations after each detection stage, i.e., $N_{\text{MP}}=\sum_{i=1}^{I}N_{i}$.
In classical IDD, this situation would correspond to $I$ IDD iterations, each of which applies $N_i = {N_{\text{MP}}}/{I}$ decoding iterations.
The solid red curve in \fref{fig:rayleigh_perf_csi_bler_performance} corresponds to a non-IDD ($I=1$) baseline with LMMSE detection followed by $12$ LDPC decoding iterations.
The solid orange curve corresponds to IDD with $I=2$.
The solid blue and solid green lines correspond to  DUIDD with LoCo-PIC and MMSE-PIC, respectively, both using $I=2$.
We see that DUIDD with MMSE-PIC performs $0.6$\,dB better than classical IDD and $2$\,dB better than the  non-iterative receiver at  $1\%$ BLER. DUIDD with LoCo-PIC loses only $0.2$\,dB at $1\%$ BLER. 

In \fref{fig:rayleigh_perf_csi_bler_duidd_vs_idd_more_dec_v2}, we analyze the detection and decoding convergence speed benefits from DUIDD. Thereby, we compare IDD with a constant number of LDPC MP decoding iterations after each detection stage ($N_i=12$), with $I=3$ DUIDD with a total number of MP iterations that remains \emph{constant} ($N_{\text{MP}}=12$, i.e., $N_i=4$).
We see that DUIDD with LoCo-PIC and fewer LDPC decoding iterations achieves approximately the same performance as conventional IDD with MMSE-PIC running for $\{I=2, N_i=12\}$ iterations. 
Even more surprisingly, $\{I=3,N_i=4\}$ DUIDD with MMSE-PIC achieves the same performance as IDD with $\{I=3,N_i=12\}$.
This highlights that DUIDD can significantly increase convergence and thereby decreases the computational complexity.

\subsection{Ray-Tracing Channels and Estimated CSI}\label{sec:sim_results_remcom}

\begin{figure}[tp]
    \centering
    \includegraphics[width=0.5\linewidth]{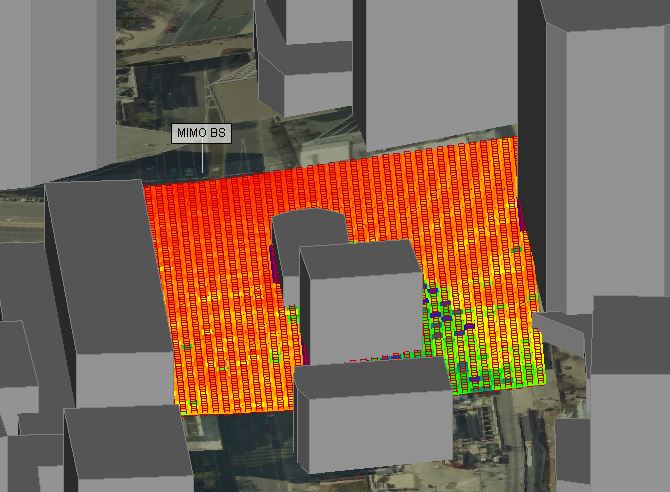}
    \caption{Visualization of the urban outdoor scenario ``Rosslyn, VA.'' generated with REMCOM's Wireless InSite ray-tracer.}
    \label{fig:remcom_scenario}
\end{figure}

\fref{fig:remcom_cest_bler_performance} shows simulation results for MIMO channel vectors generated using the REMCOM Wireless InSite ray-tracer in the urban outdoor scenario ``Rosslyn, VA.''
The scenario is depicted in \fref{fig:remcom_scenario} and uses a dual-polarized uniform linear array (ULA) at the BS with a grid of possible UE positions separated by $2.5$\,m.
For every generated MU-MIMO-OFDM channel matrix, we randomly select $U=4$ UEs from the predefined grid. 

\fref{fig:remcom_cest_bler_performance} shows simulation results where we apply a total number of $N_{\text{MP}}=12$ MP decoding iterations, which we evenly interleave with $I$ detector stages.
Remember that the total number of MP iterations equals the number of decoding iterations after each detection stage, i.e., $N_{\text{MP}}=\sum_{i=1}^{I}N_I$.
In classical IDD, this situation would correspond to $I$ IDD iterations, each of which applies $N_i = {N_{\text{MP}}}/{I}$ decoding iterations. 
The solid red curve in \fref{fig:remcom_cest_bler_performance} corresponds to a non-IDD ($I=1$) baseline with LMMSE detection followed by $12$ LDPC decoding iterations. 
The solid orange curve corresponds to conventional IDD with $I=2$.
The solid blue and solid green lines correspond to DUIDD with LoCo-PIC and MMSE-PIC, respectively, both using $I=2$.
We see that DUIDD with MMSE-PIC performs $1.2$\,dB better than conventional IDD and $2.8$\,dB better than the non-iterative receiver at $\text{BLER}=10^{-2}$. DUIDD with LoCo-PIC achieves roughly the same performance as IDD, but at lower complexity, as shown in \fref{sec:perf_vs_complexity_trade-off}.

In \fref{fig:remcom_cest_bler_duidd_vs_idd_more_dec}, we analyze the detection and decoding convergence speed benefits from DUIDD. We compare IDD with a constant number of LDPC MP decoding iterations after each detection stage ($N_i=12$), with $I=2$ DUIDD with a total number of MP iterations that remains \textit{constant} ($N_{\text{MP}}=12$, i.e., $N_i=4$). We see that DUIDD with LoCo-PIC and fewer LDPC decoding iterations achieves only slightly worse performance than IDD with MMSE-PIC running for $\{I=2,N_i=12\}$ iterations. Even more surprisingly,~$\{I=2,N_i=6\}$~DUIDD with MMSE-PIC achieves better performance than IDD with $\{I=4,N_i=12\}$.
This highlights that DUIDD can significantly improve the BLER performance and convergence speed, and thereby allows decreasing the computational complexity.

\subsection{Performance vs. Complexity Trade-Off}\label{sec:perf_vs_complexity_trade-off}

We now compare the BLER performance versus the computational complexity.
Since all of the considered methods apply the same channel estimation, noise-whitening, and LDPC decoding algorithms (with $N_{\text{MP}}=12$), we focus on the complexity of the data detector.
We count the number of real-valued multiplications $\#_{\text{MUL}}$ when implementing the detector with textbook algorithms for $T_D=10$ coherent channels.
Depending on what is optimal for each scenario and method, we consider either a Cholesky, QR, or LU decomposition for matrix inversion, and explicit calculation of the filter matrices or forward- and backward-substitution for equalization.

In Table \ref{table:complexity}, we compare the complexity of non-iterative LMMSE detection, as well as LoCo-PIC and MMSE-PIC detection, for different numbers of BS antennas $B$, UEs~$U$, and IDD iterations $I$.
We can see, compared to LoCo-PIC, the complexity of MMSE-PIC is much larger and also growing faster with $I$ and $B\times U$, because MMSE-PIC explicitly computes a matrix inverse for each coherent channel $t\in\{1,\dots,T_D\}$ and detection stage $i$~\cite{asicmimo}.
On the other hand, LoCo-PIC can reuse the filter matrix (once computed) for all $t$ and $i$, and the complexity-increase mainly results from soft-symbol computation, PIC, equalization, and LLR de-mapping.

\begin{table}[tp]
\renewcommand{\arraystretch}{1.1}
\centering
\caption{Computational complexity of data detection measured in $10^3$~real-valued multiplications for  a coherence of $T_D=10$} 
\begin{tabular}{@{}lccccc@{}}
\toprule
     & $I=1$     & $I=2$    & $I=3$    & $I=2$    & $I=3$\\
     $B\times U$           & LMMSE & LoCo- & LoCo- & MMSE- & MMSE-\\
                &  & PIC & PIC & PIC & PIC\\            
    \midrule
    $8\times4$  & $2.32$  & $4.16$        & $5.78$       & $7.47$       & $12.4$    \\
    $16\times4$ & $3.99$  & $5.76$      & $7.38$      & $9.07$      & $14.0$ \\
    $32\times16$  & $53.8$ & $81.9$        & $106$       & $255$       & $454$ \\
    \bottomrule
\end{tabular}
\label{table:complexity}
\end{table}

\begin{figure}[tp]
\centering
\subfigure[]{\includegraphics[width=.49\linewidth]{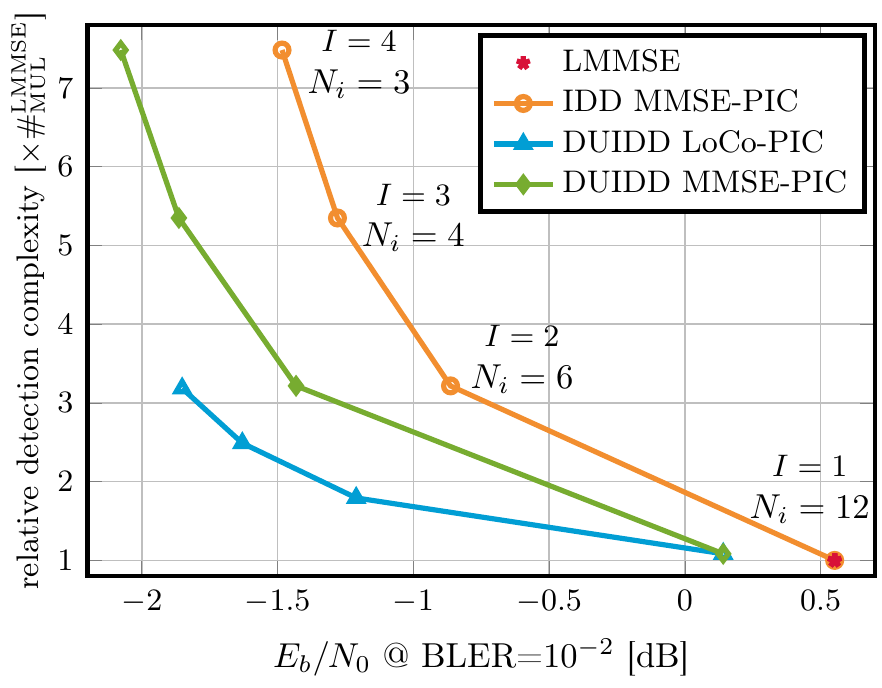}\label{fig:performance_vs_complexity_rayleigh_perf_csi}}
\hfill
\subfigure[]{\includegraphics[width=.49\linewidth]{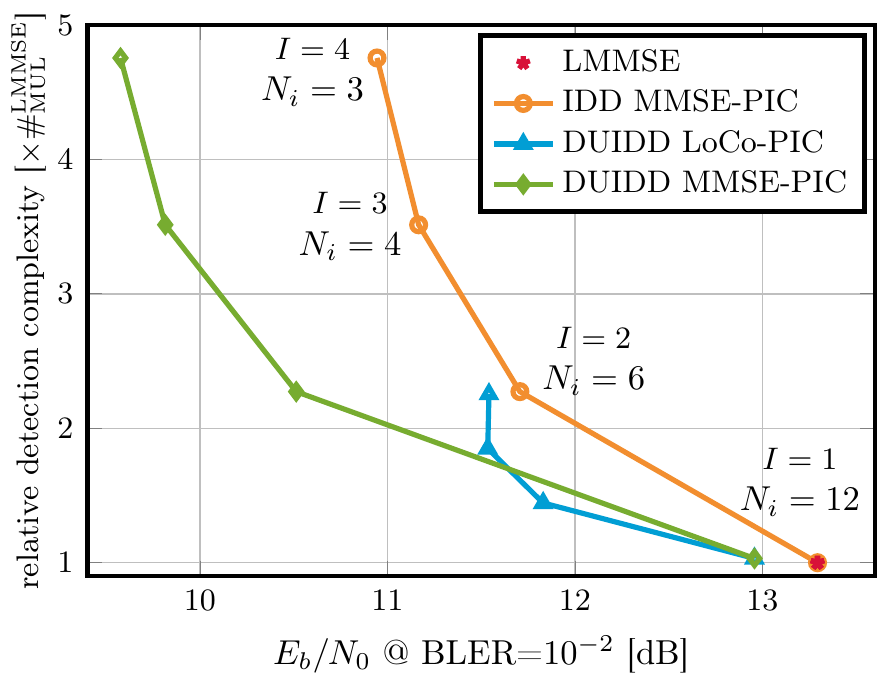}\label{fig:performance_vs_complexity_remcom_cest}}
\caption{Computational complexity vs. performance (measured in minimal SNR required to achieve $\text{BLER}=1\%$) trade-off  for a fixed number $N_{\text{MP}}=12$ of total MP decoding iterations. The subfigures consider a MU-MIMO-OFDM system (a) with $B\times U=8\times4$ i.i.d.\ Rayleigh fading channels and perfect CSI, and (b) with $B\times U=16\times4$ ray-tracing channels and estimated CSI.}
\end{figure}

\fref{fig:performance_vs_complexity_rayleigh_perf_csi} compares the SNR performance measured in the minimum $E_b/\No$ to achieve a BLER of $1$\% versus the computational complexity, normalized to the complexity of non-iterative LMMSE detection, for the Rayleigh fading scenario with perfect CSI from \fref{sec:sim_results_rayleigh}.
We see that by increasing the number of iterations $I$, the complexity increases and the SNR performance improves.
While the DUIDD MMSE-PIC receiver has the same complexity as classical IDD, DUIDD achieves superior SNR performance, with gains of up to $0.6$\,dB. In contrast, DUIDD with LoCo-PIC exhibits significantly lower complexity while achieving almost the same performance as as the MMSE-PIC-based DUIDD receiver.

\fref{fig:performance_vs_complexity_remcom_cest} compares the SNR performance versus the computational complexity for the ray-tracing scenario with estimated CSI from~\fref{sec:sim_results_remcom}.
We observe even larger SNR performance gains for DUIDD with MMSE-PIC over classical IDD, with gains of up to $1.4$\,dB. DUIDD with LoCo-PIC is only Pareto optimal for $I=2$ and outperformed by DUIDD with MMSE-PIC at higher complexity. 

% !TEX root = ../22ASILOMAR_DUIDD.tex
% DO NOT REMOVE THE ABOVE COMMENT!

\section{Conclusions}

We have proposed deep-unfolded interleaved detection and decoding (DUIDD), a novel paradigm that outperforms traditional iterative detection and decoding (IDD) receivers in terms of SNR performance and complexity.
The key idea is to break the strict separation between data detector and channel decoder in a traditional IDD receiver and to jointly optimize all algorithm parameters using deep unfolding and ML. 
To further improve the efficacy of DUIDD, we have introduced new hyperparameters that enable optimizing soft-information exchange, message damping, and state forwarding. To reduce the complexity of SISO data detection, we proposed a low-complexity variant of the SISO MMSE-PIC algorithm.
In simulations with NVIDIA's Sionna \cite{Hoydis2022}, our 5G-near MU-MIMO-OFDM DUIDD receivers
have shown SNR performance gains of up to $1.4$\,dB and reduced complexity compared to~IDD.

There are many avenues for future work. An automated search for the optimal interleaving pattern of the detector and decoder stages given a complexity constraint is ongoing. Furthermore, an investigation of other data detectors and channel decoders within the DUIDD framework may further improve the performance-complexity trade-off.

\balance

\bibliographystyle{IEEEtran}
\bibliography{bib/VIPabbrv,bib/confs-jrnls,bib/publishers,bib/VIP_190331,bib/bibliography}

% Generated by IEEEtran.bst, version: 1.14 (2015/08/26)
\begin{thebibliography}{10}
\providecommand{\url}[1]{#1}
\csname url@samestyle\endcsname
\providecommand{\newblock}{\relax}
\providecommand{\bibinfo}[2]{#2}
\providecommand{\BIBentrySTDinterwordspacing}{\spaceskip=0pt\relax}
\providecommand{\BIBentryALTinterwordstretchfactor}{4}
\providecommand{\BIBentryALTinterwordspacing}{\spaceskip=\fontdimen2\font plus
\BIBentryALTinterwordstretchfactor\fontdimen3\font minus
  \fontdimen4\font\relax}
\providecommand{\BIBforeignlanguage}[2]{{%
\expandafter\ifx\csname l@#1\endcsname\relax
\typeout{** WARNING: IEEEtran.bst: No hyphenation pattern has been}%
\typeout{** loaded for the language `#1'. Using the pattern for}%
\typeout{** the default language instead.}%
\else
\language=\csname l@#1\endcsname
\fi
#2}}
\providecommand{\BIBdecl}{\relax}
\BIBdecl

\bibitem{ETSI5G}
{European Telecommunications Standards Institute}, ``{5G} {NR} multiplexing and
  channel coding,'' Apr. 2021, {ETSI 3GPP TS 38.212 version 16.5.0 Release 16}.

\bibitem{Hoydis2021}
J.~Hoydis, F.~A. Aoudia, A.~Valcarce, and H.~Viswanathan, ``Toward a {6G}
  {AI}-native air interface,'' \emph{{IEEE} Commun. Mag.}, vol.~59, no.~5, pp.
  76--81, May 2021.

\bibitem{Seethaler2004}
D.~Seethaler, G.~Matz, and F.~Hlawatsch, ``An efficient {MMSE}-based
  demodulator for {MIMO} bit-interleaved coded modulation,'' in \emph{Proc.
  IEEE Global Telecommun. Conf. (GLOBECOM)}.\hskip 1em plus 0.5em minus
  0.4em\relax {IEEE}, Nov. 2004.

\bibitem{Burg2006}
A.~Burg, S.~Haene, D.~Perels, P.~Luethi, N.~Felber, and W.~Fichtner,
  ``Algorithm and {VLSI} architecture for linear {MMSE} detection in
  {MIMO}-{OFDM} systems,'' in \emph{IEEE Int. Symp. Circuits and Syst.
  (ISCAS)}, May 2006.

\bibitem{Ketonen2010}
J.~Ketonen, M.~Juntti, and J.~R. Cavallaro,
  ``Performance{\textemdash}complexity comparison of receivers for a {LTE}
  {MIMO}{\textendash}{OFDM} system,'' \emph{{IEEE} Trans. Signal Process.},
  vol.~58, no.~6, pp. 3360--3372, Jun. 2010.

\bibitem{WYWDCS2014}
M.~Wu, B.~Yin, G.~Wang, C.~Dick, J.~Cavallaro, and C.~Studer, ``Large-scale
  {MIMO} detection for {3GPP LTE}: Algorithms and {FPGA} implementations,''
  \emph{{IEEE} J. Sel. Topics Signal Process.}, vol.~8, no.~5, pp. 916--929,
  Oct. 2014.

\bibitem{Lu2014}
L.~Lu, G.~Y. Li, A.~L. Swindlehurst, A.~Ashikhmin, and R.~Zhang, ``An overview
  of massive {MIMO}: Benefits and challenges,'' \emph{{IEEE} J. Sel. Topics
  Signal Process.}, vol.~8, no.~5, pp. 742--758, Oct. 2014.

\bibitem{Yang2015}
S.~Yang and L.~Hanzo, ``Fifty years of {MIMO} detection: The road to
  large-scale {MIMOs},'' \emph{{IEEE} Commun. Surveys Tuts.}, vol.~17, no.~4,
  pp. 1941--1988, Sep. 2015.

\bibitem{Peng2018}
G.~Peng, L.~Liu, S.~Zhou, S.~Yin, and S.~Wei, ``A 1.58 {Gbps}/w 0.40 {Gbps}/mm2
  {ASIC} implementation of {MMSE} detection for $128\times 8~64$-{QAM} massive
  {MIMO} in 65 nm {CMOS},'' \emph{{IEEE} Trans. Circuits Syst. {I}}, vol.~65,
  no.~5, pp. 1717--1730, May 2018.

\bibitem{HT2003}
B.~Hochwald and S.~Ten~Brink, ``Achieving near-capacity on a multiple-antenna
  channel,'' \emph{{IEEE} Trans. Commun.}, vol.~51, no.~3, pp. 389--399, Mar.
  2003.

\bibitem{Studer2009diss}
C.~Studer, ``\BIBforeignlanguage{en}{Iterative {MIMO} decoding: Algorithms and
  {VLSI} implementation aspects},'' Ph.D. dissertation, {ETH} Zurich, Jul.
  2009.

\bibitem{Preyss2012}
N.~Preyss, A.~Burg, and C.~Studer, ``Layered detection and decoding in {MIMO}
  wireless systems,'' in \emph{Proc. of Conf. on Design and Archit. for Signal
  and Image Process. ({DASIP})}.\hskip 1em plus 0.5em minus 0.4em\relax IEEE,
  Oct. 2012, pp. 1--8.

\bibitem{Sun2015}
W.-C. Sun, W.-H. Wu, C.-H. Yang, and Y.-L. Ueng, ``An iterative detection and
  decoding receiver for {LDPC}-coded {MIMO} systems,'' \emph{{IEEE} Trans.
  Circuits Syst. {I}}, vol.~62, no.~10, pp. 2512--2522, Oct. 2015.

\bibitem{WDC2014}
M.~Wu, C.~Dick, J.~R. Cavallaro, and C.~Studer, ``Iterative detection and
  decoding in {3GPP} {LTE}-based massive {MIMO} systems,'' in \emph{Proc. Euro.
  Sig. Proc. Conf. (EUSIPCO)}, Sep. 2014, pp. 96--100.

\bibitem{BalatsoukasStimming2019}
A.~Balatsoukas-Stimming and C.~Studer, ``Deep unfolding for communications
  systems: A survey and some new directions,'' in \emph{{IEEE} Int'l Workshop
  on Signal Process. Sys. ({SiPS})}.\hskip 1em plus 0.5em minus 0.4em\relax
  {IEEE}, Oct. 2019.

\bibitem{Hoydis2022}
J.~Hoydis, S.~Cammerer, F.~{Ait Aoudia}, A.~Vem, N.~Binder, G.~Marcus, and
  A.~Keller, ``Sionna: An open-source library for next-generation physical
  layer research,'' \emph{arXiv preprint}, Mar. 2022.

\bibitem{Wiesmayr2022a}
R.~Wiesmayr, G.~Marti, C.~Dick, H.~Song, and C.~Studer, ``Bit error and block
  error rate training for {ML}-assisted communication,''
  \emph{arXiv:2210.14103}, Oct. 2022.

\bibitem{Tuchler2002}
M.~T\"uchler, A.~Singer, and R.~Koetter, ``Minimum mean squared error
  equalization using a priori information,'' \emph{{IEEE} Trans. Signal
  Process.}, vol.~50, no.~3, pp. 673--683, Mar. 2002.

\bibitem{asicmimo}
C.~Studer, S.~Fateh, and D.~Seethaler, ``{ASIC} implementation of soft-input
  soft-output {MIMO} detection using {MMSE} parallel interference
  cancellation,'' \emph{{IEEE} J. Solid-State Circuits}, vol.~46, no.~7, pp.
  1754--1765, Jul. 2011.

\bibitem{Nachmani2018}
E.~Nachmani, E.~Marciano, L.~Lugosch, W.~J. Gross, D.~Burshtein, and
  Y.~Be’ery, ``Deep learning methods for improved decoding of linear codes,''
  \emph{{IEEE} J. Sel. Topics Signal Process.}, vol.~12, no.~1, pp. 119--131,
  Jan. 2018.

\bibitem{Lian2018}
M.~Lian, C.~Hager, and H.~D. Pfister, ``What can machine learning teach us
  about communications?'' in \emph{Proc. IEEE Inf. Theory Workshop
  (ITW)}.\hskip 1em plus 0.5em minus 0.4em\relax {IEEE}, Nov. 2018.

\bibitem{Lin2022}
X.~Lin, ``An overview of {5G} advanced evolution in {3GPP} release 18,''
  \emph{arXiv:2201.01358}, Jan. 2022.

\bibitem{OShea2017}
T.~J. O'Shea, T.~Erpek, and T.~C. Clancy, ``Deep learning based {MIMO}
  communications,'' \emph{arXiv:1707.07980}, Jul. 2017.

\bibitem{Samuel2019}
N.~Samuel, T.~Diskin, and A.~Wiesel, ``Learning to detect,'' \emph{{IEEE}
  Trans. Signal Process.}, vol.~67, no.~10, pp. 2554--2564, May 2019.

\bibitem{Khani2020}
M.~Khani, M.~Alizadeh, J.~Hoydis, and P.~Fleming, ``Adaptive neural signal
  detection for massive {MIMO},'' \emph{{IEEE} Trans. Wireless Commun.},
  vol.~19, no.~8, pp. 5635--5648, Aug. 2020.

\bibitem{Cammerer2022}
S.~Cammerer, J.~Hoydis, F.~A. Aoudia, and A.~Keller, ``Graph neural networks
  for channel decoding,'' \emph{arXiv:2207.14742}, Jul. 2022.

\bibitem{Doerner2018}
S.~D{\"o}rner, S.~Cammerer, J.~Hoydis, and S.~ten Brink, ``Deep learning based
  communication over the air,'' \emph{{IEEE} J. Sel. Topics Signal Process.},
  vol.~12, no.~1, pp. 132--143, Feb. 2018.

\bibitem{Aoudia2019}
F.~A. Aoudia and J.~Hoydis, ``Model-free training of end-to-end communication
  systems,'' \emph{{IEEE} J. Sel. Topics Signal Process.}, vol.~37, no.~11, pp.
  2503--2516, Nov. 2019.

\bibitem{Song2022}
J.~Song, C.~H{\"a}ger, J.~Schr{\"o}der, T.~J. O’Shea, E.~Agrell, and
  H.~Wymeersch, ``Benchmarking and interpreting end-to-end learning of {MIMO}
  and multi-user communication,'' \emph{{IEEE} Trans. Wireless Commun.},
  vol.~21, no.~9, pp. 7287--7298, Sep. 2022.

\bibitem{Cammerer2020}
S.~Cammerer, F.~A. Aoudia, S.~Dorner, M.~Stark, J.~Hoydis, and S.~ten Brink,
  ``Trainable communication systems: Concepts and prototype,'' \emph{{IEEE}
  Trans. Commun.}, vol.~68, no.~9, pp. 5489--5503, Sep. 2020.

\bibitem{Zhang2022}
J.~Zhang, C.-K. Wen, and S.~Jin, ``Adaptive {MIMO} detector based on
  hypernetwork: Design, simulation, and experimental test,'' \emph{{IEEE} J.
  Sel. Areas Commun.}, vol.~40, no.~1, pp. 65--81, Jan. 2022.

\bibitem{Nguyen2022}
L.~V. Nguyen, N.~T. Nguyen, N.~H. Tran, M.~Juntti, A.~L. Swindlehurst, and
  D.~H.~N. Nguyen, ``Leveraging deep neural networks for massive {MIMO} data
  detection,'' \emph{{IEEE} Wireless Commun.}, pp. 1--7, May 2022.

\bibitem{Witzke2002}
M.~Witzke, S.~Baro, F.~Schreckenbach, and J.~Hagenauer, ``Iterative detection
  of {MIMO} signals with linear detectors,'' in \emph{Asilomar Conf. Signals,
  Syst., Comput.}\hskip 1em plus 0.5em minus 0.4em\relax {IEEE}, Nov. 2002.

\bibitem{Vogt2000}
J.~Vogt and A.~Finger, ``Improving the max-log-{MAP} turbo decoder,''
  \emph{Electron. Lett.}, vol.~36, no.~23, p.~1, Nov. 2000.

\bibitem{Tomasoni2006}
A.~Tomasoni, M.~Ferrari, D.~Gatti, F.~Osnato, and S.~Bellini, ``A low
  complexity turbo {MMSE} receiver for {W}-{LAN} {MIMO} systems,'' in
  \emph{Proc. IEEE Int. Conf. Commun. (ICC)}.\hskip 1em plus 0.5em minus
  0.4em\relax {IEEE}, Jun. 2006.

\bibitem{Collings2004}
I.~Collings, M.~Butler, and M.~McKay, ``Low complexity receiver design for
  {MIMO} bit-interleaved coded modulation,'' in \emph{Proc. {IEEE} Int'l Symp.
  Spread Spectrum Techniques and Appl.}\hskip 1em plus 0.5em minus 0.4em\relax
  {IEEE}, Aug. 2004.

\bibitem{MacKay1999}
D.~MacKay, ``Good error-correcting codes based on very sparse matrices,''
  \emph{{IEEE} Trans. Inf. Theory}, vol.~45, no.~2, pp. 399--431, Mar. 1999.

\bibitem{Savin2021}
V.~Savin, ``Gradient descent bit-flipping decoding with momentum,'' in
  \emph{Int. Symp. on Topics in Coding ({ISTC})}.\hskip 1em plus 0.5em minus
  0.4em\relax {IEEE}, Aug. 2021.

\end{thebibliography}

\end{document}